\RequirePackage{fix-cm}
\documentclass[smallextended, natbib]{svjour3} 
\smartqed  
\usepackage{paralist}
\usepackage{graphicx}
\usepackage[version = 3, upgrade = true]{acro}
\usepackage{calrsfs}    
\usepackage{caption}    
\usepackage{subcaption} 
\usepackage{hyperref}   
\usepackage{amsmath}    
\usepackage{amssymb}    
\usepackage{bm}

\usepackage{xcolor}
\usepackage{comment}
\usepackage{multirow}
\usepackage{ntheorem}

\usepackage{placeins}

\let\Oldsection\section
\renewcommand{\section}{\FloatBarrier\Oldsection}

\let\Oldsubsection\subsection
\renewcommand{\subsection}{\FloatBarrier\Oldsubsection}

\let\Oldsubsubsection\subsubsection
\renewcommand{\subsubsection}{\FloatBarrier\Oldsubsubsection}

\DeclareAcronym{ToM}{
  short = ToM ,
  long  = Theory of Mind,
  tag = abbrev
}

\DeclareAcronym{BToM}{
  short = BToM ,
  long  = Bayesian Theory of Mind,
  tag = abbrev
}

\DeclareAcronym{SAR}{
  short = SAR ,
  long  = Socially Assistive Robot, 
  tag = abbrev
}

\DeclareAcronym{POMDP}{
  short = POMDP ,
  long  = Partially Observable Markov Decision Process, 
  tag = abbrev, 
  long-plural = es,
}

\DeclareAcronym{MDP}{
  short = MDP ,
  long  = Markov Decision Process,
  tag = abbrev,
  long-plural = es,
}

\DeclareAcronym{BN}{
  short = BN ,
  long  = Bayesian Network, 
  tag = abbrev
}

\DeclareAcronym{DAG}{
    short = DAG, 
    long = Direct Acyclic Graph, 
    tag = abbrev
}

\DeclareAcronym{FCM}{
    short = FCM, 
    long = Fuzzy Cognitive Map, 
    tag = abbrev
}

\DeclareAcronym{RBFCM}{
    short = RBFCM, 
    long = Rule Based Fuzzy Cognitive Map, 
    tag = abbrev
}

\DeclareAcronym{FIS}{
    short = FIS, 
    long = Fuzzy Inference System, 
    tag = abbrev
}

\theoremseparator{:}
\newtheorem{hyp}{Hypothesis}

\usepackage[margin=3cm]{geometry}

\begin{document}

\title{Mathematical Models of Theory of Mind
}
\subtitle{Estimation and Prediction of State-of-Mind and Behaviour of Humans}


\author{Maria Mor\~{a}o Patr\'{i}cio         \and
        Anahita  Jamshidnejad*   
\vspace{-5ex}}


\institute{Both authors are with the Department of Control and Operations, Delft University of Technology, Delft, The Netherlands \\
Corresponding author email: a.jamshidnejad@tudelft.nl,   
Co-author email: marialmpatricio@hotmail.com
}

\date{}

\maketitle
\vspace{-15ex}
\begin{abstract}
Socially assistive robots provide physical and mental assistance for humans via cognitive human-machine interactions. 
These robots should sustain long-term engaging interactions with humans 
in a similar way humans interact with each other. 
According to the theory of mind, in their interactions humans develop cognitive models 
of each other in order to estimate their unobservable state-of-mind, 
predict their behavior, and act accordingly. 
Based on the theory of mind, we propose mathematical cognitive models of humans,  
which enable machines to understand cognitive procedures of humans in general 
and as distinct individuals.   
In particular, a network representation that is formulated based on  
a proposed extended version of of fuzzy cognitive maps is introduced.    
The resulting models are identified and validated using (1) 
computer-based simulations designed according to a general data set  
of human's intuitive reasoning and literature and (2) real-life personalised experiments with $15$ human participants. 
The results of the experiments show that the proposed cognitive models 
can excellently be personalised to different participants and precisely estimate and predict 
their current and future state-of-mind and expected behaviors.%

\end{abstract}

\section{Introduction}
\label{sec:intro}

Socially assistive robots (SARs), which assist humans mentally and physically via social 
interaction with them, have proven to be very successful in boosting the outcomes of 
therapeutic and educational assistance for humans 
\citep{Feil-Seifer2005,Tapus2008,Ros2019a,Scassellati2018ImprovingRobot,Kidd2008RobotsInteraction,Ros2019a,Tapus2008}.  
The majority of SAR applications, however, involve short-term interactions  
(i.e., one or very few therapeutic or educational sessions that overall last less than a month)  with humans  
that exclude in-depth analysis and understanding of every human's specific cognitive procedures 
and fixed personality, behavioral, and decision making traits \citep{Leite2013}. 
The more prominent the presence of intelligent machines, including SARs, becomes in our lives, 
the more essential becomes for machines to sustain long-term, meaningful and engaging interactions 
with humans by accounting for their personal cognitive dynamics. 
State-of-the-art SARs currently face serious challenges regarding maintaining such long-term 
engaging interactions with humans \citep{Kidd2008RobotsInteraction,Scassellati2018ImprovingRobot,Leite2013, Ros2019a}. 
Particularly, the simple rudimentary nature of social skills (including understanding and analysis of   
humans they interact with) displayed by SARs in their interactions with humans, 
significantly degrades the effectiveness and engagement level of these interactions for humans 
\citep{Leite2013, Scassellati2018ImprovingRobot, Kidd2008RobotsInteraction}. 
Although some behavioural, non-verbal cues (e.g., joint-attention\footnote{Joint-attention 
is a non-verbal skill exhibited by social agents. It implies that an agent draws the attention 
of another agent to an object by looking or pointing at the object.} \citep{Scassellati2018ImprovingRobot}, 
eye contact \citep{Kidd2008RobotsInteraction}, facial expressions \citep{Ros2019a}) 
have been implemented for SARs, due to a lack of deep cognitive understanding of humans, 
these robots often fail to recognise \emph{when} in an interaction each cue is best to be displayed \citep{Leite2013}. 
Additionally, while personalising the SAR's behavior with respect to every human \citep{Tapus2008, Ros2019a, Scassellati2018ImprovingRobot} is crucial for sustaining long-term meaningful interactions,  
due to the missing cognitive models of humans, personalisation of SARs so far has been 
task-specific and unsystematic. 
More specifically, personalisation has mainly been simplified to learning interactive 
behaviours by the SAR that maximise the score or performance of humans in 
specific therapeutic or educational tasks, rather than on explicitly learning about 
personal behaviours and characteristics of a human.
Such specific task-oriented decision making approaches assigned to SARs  
results in interactions that are perceived as less natural, attractive, and engaging for humans, especially in 
the long term.

In order to act as similarly as possible to  humans, 
SARs need to understand the dynamics of human cognition and interactions \citep{Mataric2016,Leite2013}. 
So far, the control approaches used to steer the behaviour of SARs are mostly based on 
model-free methods, e.g., reinforcement learning \citep{Tapus2008,Ros2019a}
and model-free rule-based decision making \citep{Scassellati2018ImprovingRobot,Kidd2008RobotsInteraction}. 
Ignoring the key to success in human's interactions, which according 
to the theory of mind (\ac{ToM}) \citep{Scassellati2002} is human's capability  
in (partial) modelling and awareness of the cognitive procedures and state-of-mind of 
other humans and making decisions according to such models, 
is a main bottleneck for developing intelligent machines, e.g., SARs, 
that can act and interact as closely as possible to humans. 
Therefore, this paper is focused on developing such cognitive models for 
analysis and decision making of intelligent machines.%

The rest of the paper is structured as it follows.  
\autoref{sec:previous_Work} gives an overview on previous work about proposed cognitive models of humans. 
\autoref{sec:contributions} formulates   the main contributions of the paper. 
\autoref{sec:agent-model} presents the formalisation and network representation of human-like cognitive procedures. 
\autoref{section:model_formulation} formulates the proposed cognitive models mathematically and according to 
a new extended version of fuzzy cognitive maps. 
In \autoref{sec:model_implementation} the proposed cognitive models are implemented and assessed 
via real-life experiments with $15$ volunteer human participants. 
Finally, \autoref{sec:conclusions} concludes the paper and presents topics for future research.

\section{Previous Work}
\label{sec:previous_Work}

Understanding and representing the cognitive procedures involved in human interactions 
as well as in individual cognitive dynamics of every person are of utmost importance  
for both cognitive psychology and cognitive computing (see  
\cite{Guest2021,computational_modeling_cognition_and_behavior,Mareschal2007,computational_modeling_in_cognition,Scassellati2018ImprovingRobot,Kidd2008RobotsInteraction}). 
According to the theory of mind (ToM) \cite{Scassellati2002}, 
the key to success in interactions of humans is the human's capability  
in modelling the cognitive procedures and state-of-mind of 
other humans and making decisions according to such models. 
State-of-the-art cognitive computing approaches, however, fail to maintain meaningful  
interactions with humans due to a lack of awareness of the dynamic procedures in cognition of humans 
and the personalised aspects of these procedures 
(see \cite{Feil-Seifer2005,Tapus2008,Ros2019a,Leite2013}).  
Therefore, this paper is focused on developing such cognitive models for 
analysis and decision making in cognitive computing.%


\begin{figure}
\centering
  \includegraphics[width=0.5\textwidth]{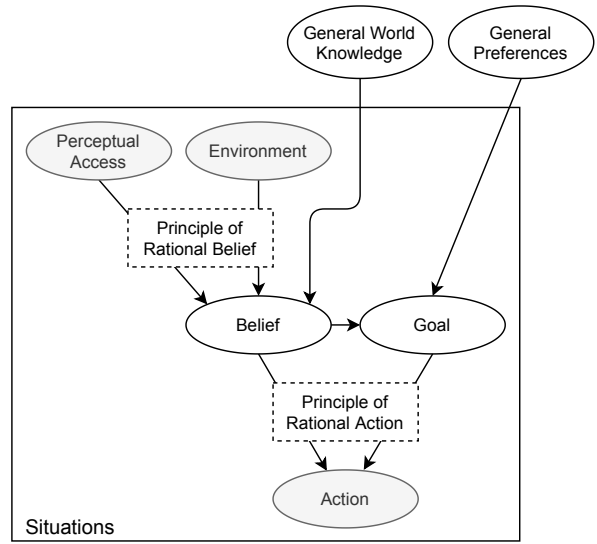}
\caption{Cognitive model proposed by Baker in \cite{Baker2012} for rational agents: 
The model explains the behaviour of rational agents based on the principles of rational action and 
and rational belief.}
\label{fig:Baker2012-PhD}       
\end{figure}

Theory of mind (ToM) has been used to develop computational frameworks based on the principles of rational belief and rational action,  
i.e., assuming that according to their \emph{rational} beliefs and goals, humans act such that they maximise their outcomes/minimise their losses \cite{Dennett1987TheStance, Baker2011, Jara-Ettinger2016ThePsychology, Saxe2017}. 
Baker in \cite{Baker2012} proposes and implements a Bayesian ToM model 
based on the network representation given in \autoref{fig:Baker2012-PhD}. 
Using partially observable Markov decision processes and Bayesian inference, the model proposed in \cite{Baker2012} 
makes forward and inverse inferences of the actions and the beliefs and goals of rational agents assuming that the  
principles of rational belief and rational action hold. 
The model was implemented in experiments where rational agents moved 
in two-dimensional spaces and the results exhibited 
close similarities with the inferences made by humans. 
These results also acknowledged that goals and beliefs 
should be inferred simultaneously as independent variables. 
However, the experiments are significantly simpler than real-life situations that involve interactions of humans. 
For instance, the model and thus the experimental scenarios in \cite{Baker2012} do not include 
the influence of varying emotions and distinct personality traits of humans in their cognitive procedures. 
Additionally, although the model in \cite{Baker2012} recognises the influence 
of general world knowledge and general preferences on beliefs and goals (see \autoref{fig:Baker2012-PhD}), 
these influences are not included in the experiments. 
Finally, the principles of rational action and rational belief are 
idealised conceptions 
that ignore personal aspects of decision making procedures of humans, 
and do not account for biases that frequently occur in their cognition. 
Although the model proposed in \cite{Baker2012} offers a highly promising framework 
for computational representation of ToM, to the best of our knowledge, there 
has been no follow-up research on extension of the model to cover more realistic and complex  
cognitive procedures of humans. 
Thus the model shown in \autoref{fig:Baker2012-PhD} is our main inspiration for proposing 
a more precise and realistic cognitive model of humans.%


\section{Main Aims and Contributions of the Paper}
\label{sec:contributions}

In this paper we develop mathematical cognitive models that estimate the state-of-mind of humans 
and predict their behavior. The main contributions of this paper include:
\begin{itemize}
    \item 
    A precise formalisation of human's cognitive procedures, based on comprehensive realistic examples,  
    represented as a network, which in addition to beliefs and goals includes dynamic 
    emotions, and systematically incorporates the influence of general world knowledge, general preferences, 
    and personality traits of rational agents. 
    These latter three elements play an important role in personalisation and precision of the resulting cognitive model.
    \item 
    Including the influence of biases, which may be boosted by emotions, personality traits, and goals, instead of proposing a universal cognitive representation based on the principle of rational belief, 
    as well as defining a new agent-specific concept called \emph{rational action selection} instead of considering 
    the universal principle of rational action. 
    \item 
    Proposing an extended version of fuzzy cognitive maps, in order to formulate mathematical models 
    for the proposed network representation within a standard state-space framework. 
    A main advantage of this mathematical representation is that the resulting cognitive models can directly 
    be embedded within various existing model-based control approaches (e.g., model predictive control) 
    to steer the decision making and interactions of cognitive intelligent machines. 
\end{itemize}
The resulting cognitive models are identified and personalised for $15$ human participants and are implemented for realistic scenarios for the participants and for simulated observed agents. 
The results of these experiments show that the proposed cognitive models precisely estimate and predict 
the current and future state-of-mind and behaviors in a personalised user-specific way. 
The proposed cognitive models will contribute to the research on \ac{ToM}, and to developing 
analyses and model-based control approaches that yield long-term engaging human-machine interactions 
and more realistic \emph{human-like} machine behaviors and machine-machine interactions.%

\section{Cognitive Models of Observed Agents: Formalisation}
\label{sec:agent-model}

In the rest of the paper, a rational\footnote{Note that although we use the term \emph{rational} agent, we mainly 
refer to cognitive procedures that follow some realistic level of rationalism that may vary from agent to agent. 
However, we do not limit our discussions to idealistic cases where rationality implies maximising the 
outcomes (or minimising the loses).} 
agent that makes inference about the state-of-mind and actions 
or behaviours of another rational agent is called an \emph{observer agent}. 
The other rational agent is called an \emph{observed agent}. 
In the following discussions we formalise cognitive models that describe how an observer agent 
makes inference about an observed agent.%

The state-of-mind and thus actions and behaviours of an observed agent are influenced by its 
fixed (or more precisely invariant in longer terms) characteristics including 
the agent's general world knowledge, general preferences, and personality traits, 
as well as by its dynamic (or more precisely varying more frequently in time) inner state variables 
including beliefs, goals, and emotions. 
Moreover, the environmental data of an observed agent is perceived by the agent in a personalised way. 
Next we explain how the proposed cognitive models incorporate these characteristics.%

\subsection{Variables and Dynamics of Cognitive Models}
\label{subsec:variables-description}

We propose a network representation (see \autoref{fig:model3-with-emotions}) of human's cognitive procedures that is 
composed of elements connected via directed links, which represent the inter-dependencies  
and influences of these elements. 
\autoref{app:examples} gives several representative examples based on real-life 
scenarios used to develop the network representation, which will be discussed in detail. 
The main elements of the network shown in \autoref{fig:model3-with-emotions} correspond to: 
\begin{enumerate}
    \item 
    External (uncontrolled) inputs to the cognitive model, such as   
    environmental factors (e.g., weather conditions) that may influence 
    the state-of-mind and thus actions and behaviours of observed agents. 
    \item
    State variables of the cognitive model, including the state-of-mind variables 
(beliefs, goals, emotions) of an observed agent. 
    \item
    Fixed parameters of the cognitive model, including general world knowledge, general preferences, and 
    personality traits of observed agents.
    \item
    Dynamic processes, which are functions that receive the fixed parameters and current external inputs and state variables
    and update the next-step state variables of the observed agent or predict its behaviours. 
\end{enumerate}

\subsubsection{Elements internal and external to an observed agent}

\begin{figure}
\centering
  \includegraphics[width=\textwidth]{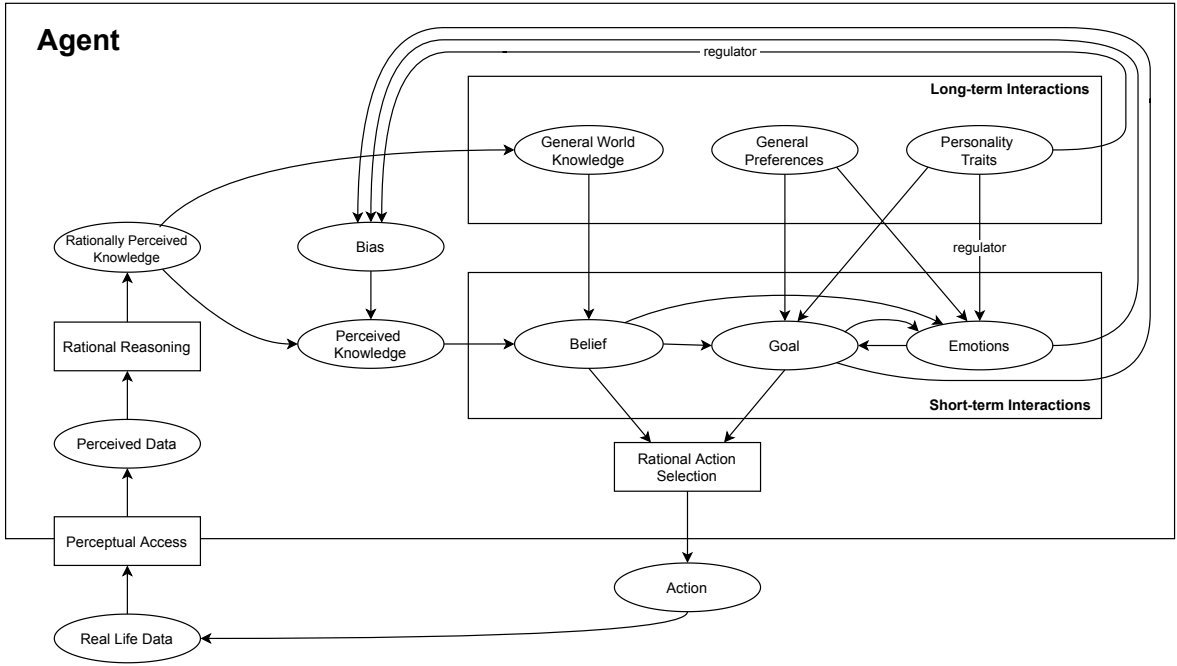}
    \caption{Proposed network representation of human's cognitive procedures including emotions, personality traits, and biases. 
    Oval-shaped elements show (input, output, and state) variables and rectangular elements correspond to processes or functions.}
\label{fig:model3-with-emotions}       
\end{figure}

While \cite{Baker2012} distinguishes the elements in the cognitive network representation 
of \autoref{fig:Baker2012-PhD} based on whether or not they depend on situations,  
we differentiate the elements of our proposed cognitive network representation 
(\autoref{fig:model3-with-emotions}) considering whether or not they are external to an observed agent. 
The main advantages of our proposed approach include: 
\begin{itemize}
    \item 
    The developed mathematical cognitive models correspond to 
    the existing standard (e.g., state space) frameworks of mathematical modelling and systems theory, allowing 
    to directly use them with various model-based decision making approaches. 
    \item
    This categorisation allows to personalise the rationalisation and action selection per observed agent.  
    \item
     Since the internal elements of an observed agent  
    are not observable for the observer agent, their inference is in general personalised to an observer agent  
    (although out of the scope of this paper, our framework is then easily expandable for 
    cases where second-order inferences, 
    i.e., inference about the inference of an observer agent about the observed agent \cite{Baker2008},  
    are of interest.  
\end{itemize}
In addition to fully internal (i.e., unobservable) and fully external 
(i.e., observable) elements, our representation can incorporate partially external (i.e., partially observable) elements. 
In particular, the perceptual access of an observed agent is partially external, given that this process is influenced by external inputs, and is partially internal, since it is shaped by internal characteristics of the individual. This dual influence on perceptual access is explained in detail in \autoref{subsec:observation-reasoning}.

    
\subsubsection{Fast-dynamics and slow-dynamics state variables} 

The state variables of the proposed cognitive model are distinguished according to their 
relevance for duration (i.e., short-term or long-term) of interactions between two rational agents and 
to the frequency of their dynamics. Consequently, two categories of state variables are defined: 
(i) Fast-dynamics state variables, which may constantly vary 
(with a time scale in the range of seconds or minutes) as a response to specific situations the observed agent faces. 
Goals, beliefs, and emotions in \autoref{fig:model3-with-emotions} are fast-dynamics state variables.
(ii) Slow-dynamics state variables, which vary according to large time scales (months or years). 
General world knowledge, general preferences, and personality traits in \autoref{fig:model3-with-emotions} 
are slow-dynamics state variables. 
Fast-dynamics state variables are more relevant for short-term interactions, while slow-dynamics state variables become more relevant throughout long-term interactions, 
when the fixed or repetitive patterns of cognitive procedures  
resulting from these slow-dynamics state variables provide extra information 
for the observer agent to make more precise estimates and predictions.%

Goals are immediate desires and needs of rational agents, such as finding food or reaching a location. 
General preferences of rational agents build up in long terms and remain invariant for long, 
and in order to identify them several interactions with a rational agent is needed. 
Examples of general preferences include favourite tastes, fiends, and hobbies of a rational agent. 
Beliefs correspond to \emph{temporary} knowledge or interpretations of rational agents from their world, 
while general world knowledge consists in \emph{persistent} rationally perceived knowledge, which 
remains unchanged or is rarely updated. 
For example, a rational agent believes that a friend who has left an hour ago to fetch 
a medicine from the drugstore is now in the city center, 
whereas the exact location of the drugstore is the agent's general world knowledge. Next we explain these elements and their mutual influences with respect to the 
other elements in the proposed cognitive network representation.%

\begin{remark}
\label{remark:slow-dynamics-not-changed}
Slow-dynamics state variables may influence the evolution of fast-dynamics state variables 
(see \autoref{example:belief_guess}-\autoref{example:belief_personalitytraits} in \autoref{app:examples}), 
while the opposite is not necessarily true (especially in short terms). 
The main aim of this paper is to formalise and formulate the evolution of fast-dynamics state variables. 
Modelling the evolution of slow-dynamics state variables is out of the scope of this paper. 
Thus slow-dynamics state variables are mainly considered as fixed parameters in the proposed cognitive models.
\end{remark}


\subsubsection{Emotions and personality traits}
\label{subsec:emotions-and-personality-traits}

In interactions between rational agents identifying the emotions of observed agents is of utmost importance  
to enable observer agents to make an overall inference about the state-of-mind of observed agents \citep{Kwon2008, Saxe2017} 
and to select accordingly the most appropriate behaviour \citep{Tapus2008,Zaki2013, Lee2019}.  
Moreover, the personality traits of observed agents act as regulators of their emotions \citep{Bono2007PersonalitySelf-monitoring}. 
Therefore, by incorporating both the emotions and personality traits in cognitive models 
more genuine, engaging, and human-like interactions will be possible for machines that use these models \citep{Tapus2008,Leite2013}.%

Personality traits are slow-dynamics state variables, therefore we are interested in 
understanding their potential influences on the fast-dynamics state variables. 
Beliefs may (indirectly) be affected by personality traits through generated biases. 
Goals may directly be affected by personality traits (see \autoref{fig:model3-with-emotions}): 
for instance, while the goal of an introvert rational agent is to avoid strangers, 
the goal of an extrovert rational agent is to make new friends. 
Finally, emotions are influenced directly by the personality traits, where this 
influence together with the potential influences of other state variables are discussed next.%

\paragraph{State variables that influence emotions:} 
Emotions of a rational agent may be stimulated by its beliefs that are generated by   
the rationally perceived knowledge (see \autoref{fig:model3-with-emotions} and 
\autoref{example:belief_generating_emotions} in \autoref{app:examples}). 
Note that emotions are not directly generated by external inputs (i.e., real-life data received by 
a rational agent), but by how the rational agent internalises and interprets these inputs.

Both goals and general preferences, alongside a belief, can impact emotions (see \autoref{fig:model3-with-emotions}). 
On the one hand, when a rational agent follows a goal and develops a belief that is in line with the fulfilment of that goal, positive emotions may be stimulated. 
On the other hand, when a rational agent follows a goal and develops a belief that 
hinders the chances of fulfilling that  goal, negative emotions may be stimulated.  
Similarly when a general preference is supported by a developed belief 
positive emotions can be generated, while beliefs that conflict with the general preferences 
may result in  negative emotions (see \autoref{example:goal_to_emotion}  in \autoref{app:examples}). 
General preferences mostly influence the emotions of a rational agent indirectly and via 
generating a goal - that alongside a belief - stimulates emotions 
(see \autoref{example:general_preference_to_emotions} in \autoref{app:examples}),  
but the influence is in some cases direct (see \autoref{fig:model3-with-emotions}). 
Any direct influence from the general world knowledge on the emotions is negligible.     
In practice, general world knowledge is transformed into beliefs, which influence the emotions as discussed before.%

Personality traits of rational agents may determine the extent to which 
certain beliefs affect their emotions: for instance, 
while both an introvert and an extrovert rational agent become happy for receiving a birthday present, 
the extrovert one experiences more excitement.  
Personality traits do not generate emotions by themselves, which is in line with the fact that emotions are fast-dynamics state variables that are temporary and event-triggered, 
whereas personality traits constantly exist. 
In other words, if personality traits were directly generating emotions a rational agent had to continuously experience those emotions. Personality traits instead boost or hinder the emotions and may be seen as regulators of the emotions.%

\begin{remark}
\label{remark:emotion-triggers}
In summary, beliefs (either alone or supported by generated goals or by general preferences)  
and boosted or hindered by personality traits trigger the emotions.
For the sake of brevity of the notations, we use the following terminology: 
\textbf{emotion trigger 1} for a combination of beliefs and general preferences, 
\textbf{emotion trigger 2} for solely beliefs, and \textbf{emotion trigger 3} for a combination of beliefs 
and goals.%
\end{remark}






\paragraph{State variables that are influenced by emotions:}  
Studies show that emotions can affect the immediate goals and desires\footnote{In this paper, 
the concepts of goals, desires, wishes, and needs of a rational agent are used interchangeably.} 
of rational agents  \citep{Raghunathan1999, Andrade2009, Lerner2015, George2016}. 
More specifically, emotions may result in the development of a goal that contradicts the general preferences of a rational agent or in the change of a goal that was previously made by the rational agent.  
For instance, gratitude  can galvanise rational agents into helping others \citep{Lerner2015}, 
or anxiety may trigger rational agents to avoid stressful situations \citep{Raghunathan1999}. 
The influence of emotions over goals is introduced into the proposed cognitive model 
shown in \autoref{fig:model3-with-emotions} via a directed link 
(also see \autoref{example:emotions_to_change_goals} in \autoref{app:examples}).

While emotions do not directly influence beliefs, they can affect the processes that result in judgements or beliefs 
of rational agents \citep{Raghunathan1999, Andrade2009}. 
More specifically, positive emotions may introduce optimistic biases into the process of generation of new beliefs, 
whereas negative emotions may lead to the formation of overly pessimistic beliefs \citep{Lerner2015}. 
Therefore, the influence of emotions on the development of beliefs will be introduced into the proposed cognitive model.%


\subsubsection{Perceptual access and rational reasoning}
\label{subsec:observation-reasoning}

General world knowledge and beliefs are acquired by rational agents through the same procedures. 
In the model proposed in \cite{Baker2012} these procedures are represented as a single element called the principle of rational belief (see \autoref{fig:Baker2012-PhD}). 
This simplification was shown to be sufficient to explain the relationship between the environmental inputs and the inferred beliefs in the simple environments and scenarios considered in \cite{Baker2012}. 
In real-life scenarios, however, a more complicated procedure occurs before a belief or a piece of general world knowledge 
is developed based on the raw real-life data. 
Data that is perceived by rational agents may differ from the real-life data. 
More specifically, rational agents may deliberately or indeliberately 
 access and perceive only a portion of the real-life data in each interaction with their environment.  
On the one hand, perception depends on rational agents, i.e., when located in the same environment 
different rational agents may notice different types of data (e.g., one may perceive a sound that is inadvertently 
heard while another agent filters the sound out).
On the other hand, rational agents may receive only partial real-life data due to 
external factors (e.g., missing visual data due to occlusion). 
Therefore, rational agents may hold false or inaccurate beliefs (c.f.\ the Sally-Anne experiment \cite{Baron-Cohen1985Does}), 
which is essential for observer agents to recognise \cite{Wellman2001Meta-analysisBelief,Rabinowitz2018}.%

To address these aspects, the process that transforms real-life data into beliefs and general world knowledge 
is decomposed into smaller, well-defined sub-processes, \emph{perceptual access} and \emph{rational reasoning} (see \autoref{fig:model3-with-emotions}). 
Real-life data from the environment is perceived via \emph{perceptual access},  
which, as explained above, is partially personalised and partially depends on the environment. 
Thus the corresponding rectangular element in \autoref{fig:model3-with-emotions}) 
is located at the border of the agent box. 
The perceived data is then processed via \emph{rational reasoning}, which as opposed to the principle 
of rational reasoning applied in \cite{Baker2012}, is specific to a rational agent.  
Accordingly the rational agent makes a judgement, i.e., the \emph{rationally perceived knowledge} (see \autoref{fig:model3-with-emotions}),  
which may be transformed into a belief or a general world knowledge by the rational agent. 
In particular, \autoref{example:personalized_perception} and \autoref{example:decomposing_rational_perception} 
in \autoref{app:examples} illustrate the importance of personalisation of the perception procedure 
and decomposing the procedure of developing rationally perceived knowledge into the proposed sub-procedures.%



\subsection{Inverse Inference of Emotions from Actions}
\label{subsec:emotion-inference-from-actions}

\begin{figure}
    \centering
    \includegraphics[width=\textwidth]{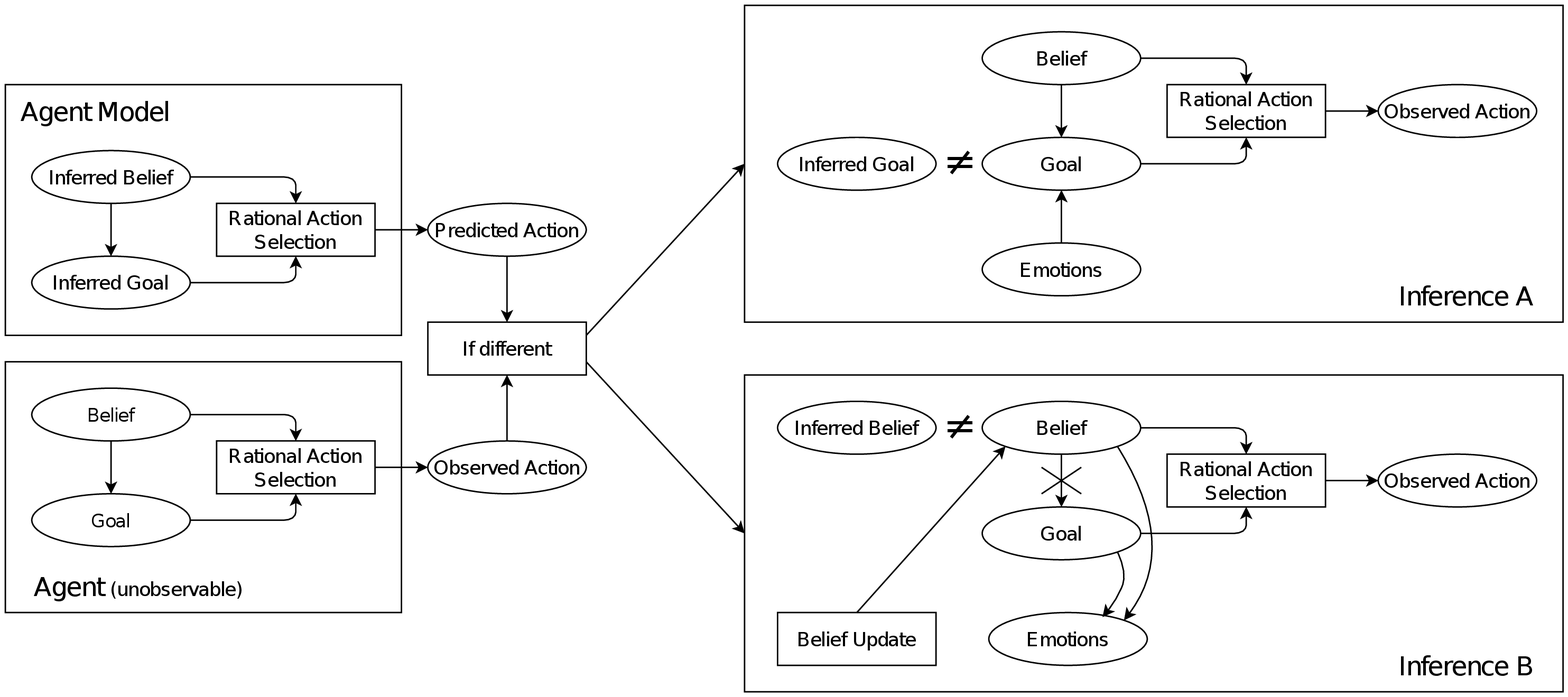}
    \caption{Inference of emotions from actions: Whenever emotions exist, the observed actions of a rational agent may differ 
    from those that are predicted based on cognitive models that include beliefs and goals only (left-hand side plots). 
    In such cases, the agent's goal or belief (top right and bottom right plots respectively) 
    or both have been influenced by the agent's emotions.}
    \label{fig:emotion-inference}
\end{figure}

According to the principle of rational action, 
actions of rational agents are a direct consequence of their beliefs and goals 
\cite{Dennett1987TheStance, Baker2011, Jara-Ettinger2016ThePsychology, Saxe2017}. 
Authors in \cite{Baker2007, Baker2011} discussed inverse inference of the beliefs and goals from actions.  
By introducing two new elements, emotion and bias, and by changing the universal 
principle of rational action to rational action selection in our proposed cognitive models, 
we should take into account the influence of these elements on the observed actions 
and also inverse inference of these elements based on the observed actions.  %
Contrarily to beliefs and goals, emotions do not directly generate actions, while  
they contribute to the generation of goals and beliefs     
and thus indirectly to the observed actions of rational agents. 
In particular, estimation of the emotions (and thus biases) based on observed actions is done via  
our proposed model next to a simplified version of this model that considers the belief-goal pair only 
(see \autoref{fig:emotion-inference}).  
Observing an action different from the action predicted by such a model implies that either the belief or the goal or both 
have been different from those considered by the cognitive model, meaning that the belief 
(see \autoref{remark:emotion-triggers}) 
corresponding to one or more simulation steps earlier 
has triggered emotions that have resulted in different goals and beliefs for the current simulation step than those estimated 
by the simplified model 
(see \autoref{example:inverse_inference_belief} and \autoref{example:inverse_inference_goal} in \autoref{app:examples}).  
Thus for inverse inference of the emotions, knowing or inferring the beliefs and goals of an observed agent 
for at least two simulation steps is needed. 
A forward inference of the belief-goal pair 
leads to the prediction of an expected action. 
Next, comparing the observed and predicted actions of the observed agent, 
the underlying emotions triggered in the previous simulation steps and influencing the belief, goal, or both 
in the current simulation step are inferred. 
Such combination of forward and inverse inferences was previously mentioned by \cite{Saxe2017}, 
although no specific framework for implementing it was proposed or discussed. 
Note that the estimations obtained via a forward inference 
may be analysed together with the estimations obtained via an inverse inference from the observed actions 
within two parallel computation modules to provide more accurate estimations and predictions.%

\section{Cognitive Models of Observed Agents: Formulation}
\label{section:model_formulation}

\begin{figure}
    \centering
    \includegraphics[width=0.65\textwidth]{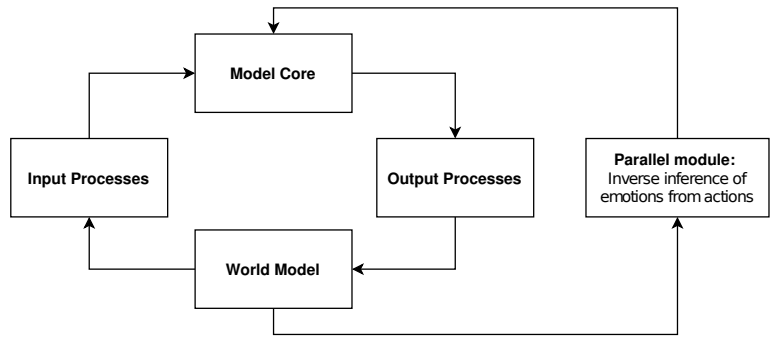}
    \caption{Different modules of the proposed modelling framework: 
    \textbf{Model core} includes the \textit{internal variables} that influence the fast-dynamics state variables. 
    \textbf{Input processes} include the perceptual access and rational reasoning. 
    \textbf{Output processes} include the rational action selection block. 
    \textbf{World model} formulates the  influence of the agent's actions on real-life data. 
    Inverse inference of emotions from actions is represented in a module parallel to the model core.}
    \label{fig:model_formulation-division}
\end{figure}



In order to facilitate the formulation of the proposed cognitive network representation, it has been 
broken into five sub-modules (see \autoref{fig:model_formulation-division}): 
(1) \emph{model core}, including the internal variables that play a role in the dynamic 
evolution of the fast-dynamics state variables, 
(2) \emph{input processes}, including the perceptual access and rational reasoning functions and auxiliary variable perceived data,  
(3) \emph{output processes}, including rational action selection,  
(4) \emph{world model}, which formulates the influence of the rational agent's actions on real-life data, and 
(5) \emph{parallel inverse inference module}, which inversely infers the emotions of the rational agent according to the observed actions 
as discussed in \autoref{subsec:emotion-inference-from-actions}.%
 
\begin{figure}
    \centering
    \includegraphics[width=.72\textwidth]{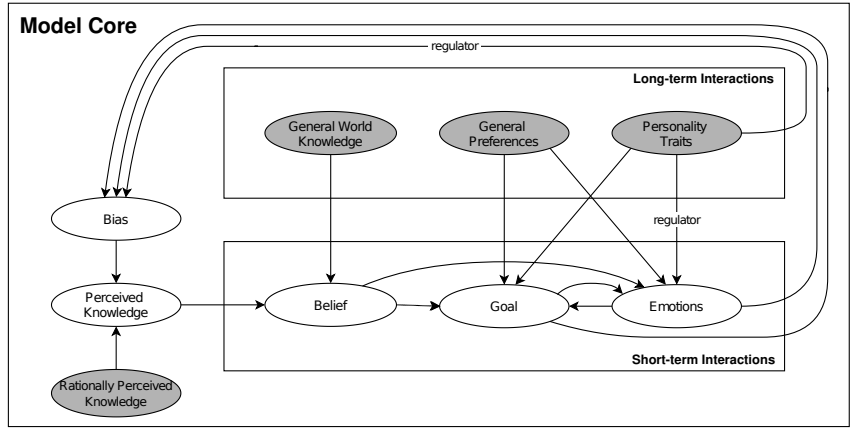}
    \caption{Model core: Elements that are state and auxiliary variables for the model core are represented in white, 
    while elements that are inputs of the model core are represented in grey.}
    \label{fig:model_modules}
\end{figure}

Since the main aim of this paper is modelling the dynamic evolution of fast-dynamics state variables, 
we focus on mathematical formulation of the model core. 
\autoref{fig:model_modules} shows the elements of the model core, where the model's state variables (i.e., belief, goal, and emotion) and auxiliary variables (perceived knowledge and bias) are represented in white and the model's inputs and fixed parameters (i.e., rationally perceived knowledge, general world knowledge, general preferences, and personality traits) are illustrated in grey.%



Previous works \citep{Baker2011,Baker2017,Jara-Ettinger2016ThePsychology,Saxe2017,Lee2019} 
mainly use Bayes' theorem to describe human's cognition. This basically implies that humans make cognition in terms of probabilities  
by developing connections between different premises and by inferring about the likelihood of various random events based on their prior knowledge \citep{Johnson-Laird1994MentalThinking}. 
Using these assumptions corresponds to considering the cognitive network representation as a Bayesian network \citep{Heckerman2008}. 
Alternatively, fuzzy cognitive maps (FCMs) \citep{FCM:BartKosko} represent concepts and variables that correspond to complex and/or uncertain systems and their interlinks and interactions. 
Contrarily to Bayesian networks, FCMs support cyclic connections \citep{Stylios2004ModelingMaps}, which is very relevant for 
modelling cognitive procedures of humans (c.f.\ \autoref{fig:model_modules}). 
Moreover, concepts or variables in an FCM can be represented mathematically by fuzzy variables, which excellently fit 
those concepts that are involved in human cognitive procedures (e.g., beliefs, goals, emotions, etc.). 
Therefore, in this paper, based on the idea of FCMs, we propose an extended FCM representation and use it to mathematically formulate 
our proposed cognitive models.%


\begin{figure}
    \centering
    \begin{subfigure}{0.4\textwidth}
        \centering
        \includegraphics[width=\textwidth]{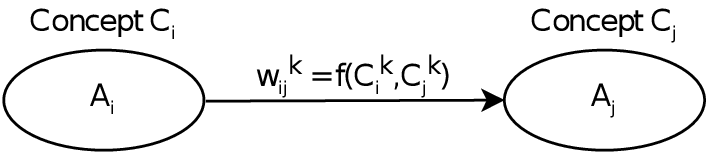}
        \caption{Simple linkage}
        \label{fig:FCM_simple_linkage}
    \end{subfigure}
    \hspace{5ex}
    \begin{subfigure}{0.45\textwidth}
        \centering
        \includegraphics[width=\textwidth]{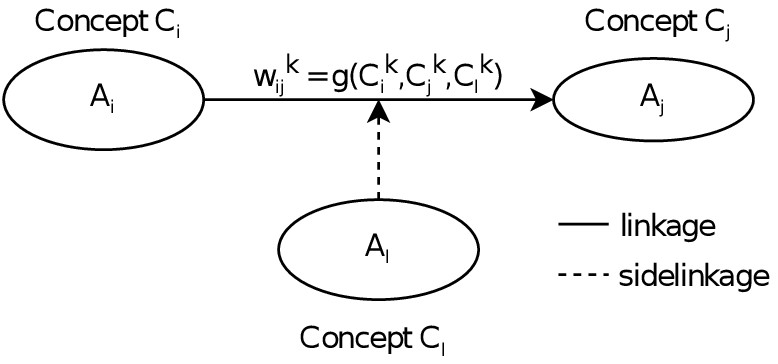}
        \caption{Complex linkage}
        \label{fig:FCM_complex_linkage}
    \end{subfigure}
    \caption{Simple and complex linkage}
    \label{FCM:linkage&sidelinkage}
\end{figure}

In an FCM the elements or variables that explain the evolution of the system are called \textit{concepts}. 
The $i^{\textrm{th}}$ concept of the system is denoted by $C_i$. 
Mathematically, $C_i$ (e.g., emotion) can be represented as a fuzzy variable, with $A_i$ a possible representation (e.g., happy) of it.  
We define $\mathbb{C}$ as the set of all concepts and $\mathbb{A}$ as the set of all possible realisations of these concepts. 
The directed influence of concept $C_i$ over concept $C_j$ in an FCM is represented by a directed line called a \textit{linkage} (see \autoref{fig:FCM_simple_linkage}).
Every linkage is characterised by a weight $w_{ij} \in [-1,1]$ that reflects the level of influence of concept $C_i$ over concept $C_j$. 
Whenever $w_{ij}$ is positive (negative), an increase in realisation $A_i$ of $C_i$ implies an increase (a decrease) in realisation $A_j$ of $C_j$ (the larger the absolute value of $w_{ij}$, the larger the influence of $C_i$ over $C_j$). 
Whenever $w_{ij}$  is null, changes in realisation $A_i$ of $C_i$ do not influence realisation $A_j$ of $C_j$.%

In FCMs the weights $w_{ij}$ are considered to be constant. 
However, in order to accurately model most real-world systems with FCMs variable weights may be required   \citep{Carvalho2001RuleDynamics,Mourhir2016AAssessment}. 
For instance, in rule-based FCM \citep{Carvalho2001RuleDynamics} the values of weights depend on the realised value $A_i$ of 
the causing variable or concept $C_i$. 
In our proposed cognitive network representation, in some cases the value of weight $w_{ij}$ for a given simulation step $k$ may 
depend on the realised values $A_i$, $A_j$, or $A_\ell$ of the causing concept $C_i$, affected concept $C_j$, or another 
intermediate concept $C_\ell$ corresponding to that simulation step. 
To address these requirements, we consider weights that may be a function of the causing, affected, or intermediate concepts 
and accordingly define \emph{simple linkages}, \emph{side linkages}, and \emph{complex linkages} and introduce their mathematical representations.%

The linkages $(i,j)$ that connect two concepts $C_i$ and $C_j$ directly and are not influenced by an intermediate concept (see \autoref{fig:FCM_simple_linkage} 
and \autoref{fig:FCM_simple_linkage}) are called \textit{simple linkages}.
A \textit{side linkage }$(\ell,i,j)$ corresponds to the directed influence of an intermediate concept $C_\ell$ 
over a linkage $(i,j)$ that connects concepts $C_i$ and $C_j$ (see the dashed arrow in \autoref{fig:FCM_complex_linkage}). 
The collection of a linkage that is influenced by one or several side linkages and all those side linkages 
is called a \textit{complex linkage} (see \autoref{fig:FCM_complex_linkage}). 
The set of all ordered pairs $(i,j)$ corresponding to simple linkages is given by $\mathbb{L}$ and the set of all 
ordered trios\footnote{In our cognitive network representation, complex linkages include no more than one side linkage.} 
$(\ell,i,j)$ that correspond to complex linkages is given by $\overline{\mathbb{L}}$.
The weight of a simple and a complex linkage for simulation step $k$ is computed via, respectively,  
function $f:\mathbb{A}^2 \rightarrow [-1,1]$ and function $g:\mathbb{A}^3 \rightarrow [-1,1]$. we have:
\begin{align}
\begin{array}{ll}
    w_{ij} (k) = f \left( C_i(k) , C_j(k)\right), \qquad
    & \forall i,j \quad \textrm{for which}\quad  (i,j) \in \mathbb{L}\\
    w_{ij} (k) = g \left( C_\ell(k) , C_i(k), C_j(k)\right),\qquad
    & \forall i,j \quad \textrm{for which}\quad \exists \ell \quad \textrm{such that} \quad   (\ell,i,j) \in \overline{\mathbb{L}}
\end{array}
\end{align}
for all $k\in\{1,2,\ldots\}$, $C_\ell,C_i,C_j \in \mathbb{C}$, and $C_\ell(k), C_i(k), C_j(k)\in \mathbb{A}$.%

 
The dynamic equation for updating a concept $C_j\in\mathbb{C}$ that evolves per simulation step $k$ within the proposed extended FCM is formulated by:
\begin{align}
    \label{eq:update-FCM}
    C_j(k+1) = h
    \Biggl(
    \sum_{\forall i | (i,j) \in \mathbb{L}} f\Big(C_i(k),C_j(k)\Big) C_i(k)     
    + 
    \sum_{    \forall i ;\exists \ell | (i,j,\ell) \in \overline{\mathbb{L}}}
    g\Big(C_\ell(k),C_i(k),C_j(k)\Big) C_i(k) 
    + 
    \alpha_{j} C_j(k)
    \Biggr)
\end{align}
where $h(\cdot)$ is in general a threshold function that constrains the evolving concept $C_j$ to remain within 
its admissible set $\mathbb{A}_j \subseteq \mathbb{A}$ and $\alpha_j$ determines the influence of the realised value 
of concept $C_j$ for simulation step $k$ on its value for simulation step $k+1$. 
Note that for the proposed cognitive model, concept $C_j$ in \eqref{eq:update-FCM} should correspond to one of the 
fast-dynamics state variables, while concept $C_i$ may be another fast-dynamics state variable, an auxiliary variable, a 
slow-dynamics state variable, or an input variable (see \autoref{fig:model_modules}).%

In order to define the functions $h(\cdot)$, $f(\cdot)$, and $g(\cdot)$ for the FCM corresponding to the proposed cognitive models, 
different approaches may be used, such as using crisp mathematical functions or describing them via 
fuzzy inference systems (see, e.g., \citep{Carvalho2001RuleDynamics}).%

\section{Cognitive Models of Observed Agents: Implementation}
\label{sec:model_implementation}

The proposed cognitive model was implemented via MATLAB and was used to simulate the expected cognitive procedures of 
human-like computer-based rational agents and human participants in several real-life scenarios. 
For qualitative assessment of the cognitive model (i.e., for 
validating the trends of the dynamic evolution of the fast-dynamics state variables beliefs, goals, and emotions), 
computer-based rational agents simulated according to  
a generalised knowledge base (built upon intuitive data provided by human participants 
and literature) are used.     
Next, $15$ human participants were asked to fill in online surveys, where they 
provided personalised answers regarding their state-of-mind for several real-life scenarios.  
The developed cognitive model was used to estimate state-of-mind of the participants within the same scenarios. 
These estimations were compared to the answers directly provided by the participants to assess the models.

\subsubsection*{Implementation Setup}
\label{subsec:implementation-setup}

\begin{figure}
\centering
    \begin{subfigure}{\textwidth}
        \centering
        \includegraphics[width=\textwidth]{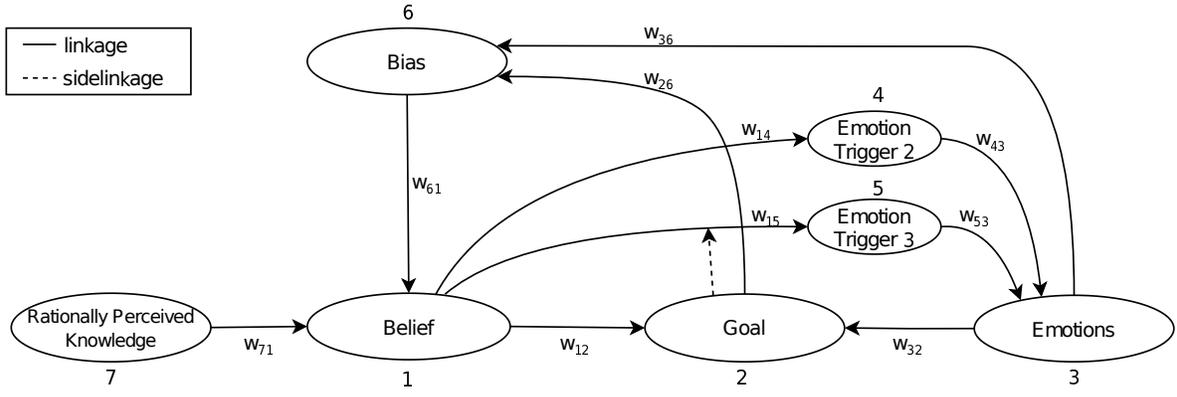}
        \caption{Network representation for \textit{real-life situation 1} illustrating auxiliary variables (emotion triggers 2 and 3) and excluding slow-rate state variables (general world knowledge, general preferences, personality traits).}
        \label{fig:situation1}
    \end{subfigure}
    \begin{subfigure}{\textwidth}
        \centering
        \includegraphics[width=\textwidth]{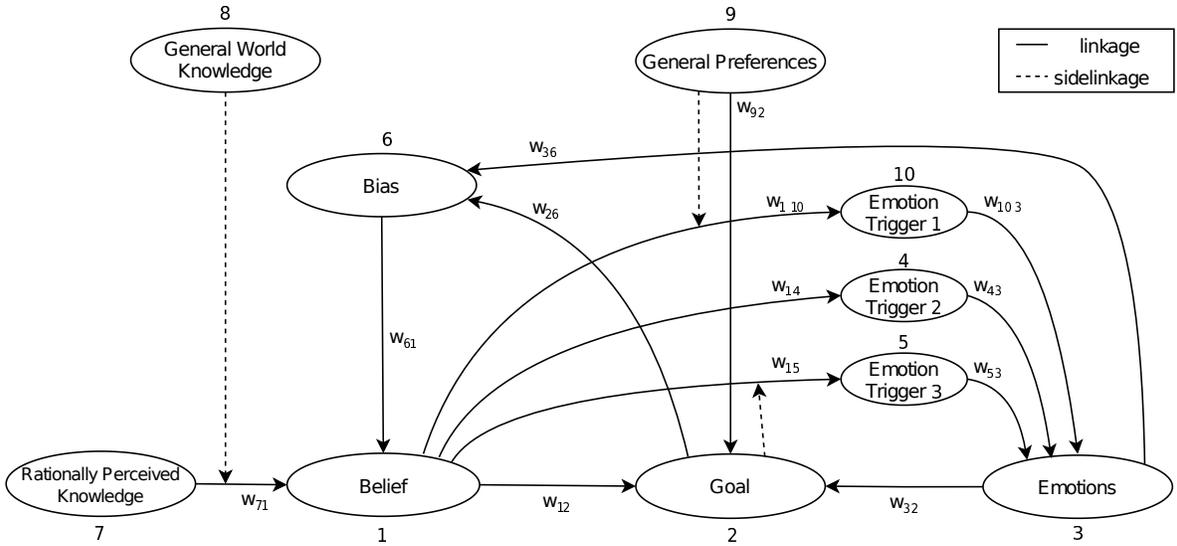}
        \caption{Network representation for \textit{real-life situation 2} illustrating auxiliary variables (emotion triggers 1, 2, and 3).}
        \label{fig:situation2}
    \end{subfigure}
    \caption{Proposed network representation of human's cognitive procedures: 
    Oval-shaped and rectangular elements show, respectively, (input, output, and state) variables and processes/functions.
    }
    \label{FCM:real-life-situations}
\end{figure}

\begin{table}
\centering
\begin{tabular}{|l|l|lll|l}
\hline
\multicolumn{1}{|c|}{\multirow{2}{*}{\textbf{Concept}}} & \multicolumn{1}{c|}{\multirow{2}{*}{\textbf{Scenario}}} & \multicolumn{3}{c|}{\textbf{Linguistic term}}  \multirow{2}{*}{\textbf{\begin{tabular}[c]{@{}l@{}}\end{tabular}}} \\ \cline{3-5}
\multicolumn{1}{|c|}{} & \multicolumn{1}{c|}{} & \multicolumn{1}{c|}{\textbf{Min realisation $-1$}}  
& \multicolumn{1}{c|}{\textbf{Median realisation $0$}} 
& \multicolumn{1}{c|}{\textbf{Max realisation $1$}}   \\ \hline
\textbf{Belief} & 1, 2  & \multicolumn{1}{l|}{\begin{tabular}[c]{@{}l@{}}There will be heavy rain\end{tabular}} & \multicolumn{1}{l|}{\begin{tabular}[c]{@{}l@{}} No information about the weather\end{tabular}} & It will be very sunny\\ \hline
\textbf{Goal} & 1, 2 & \multicolumn{1}{l|}{\begin{tabular}[c]{@{}l@{}}Agent does not want to do \\ the outdoor activity\end{tabular}} & \multicolumn{1}{l|}{\begin{tabular}[c]{@{}l@{}}Agent does not have a  preference  \\ about the outdoor activity\end{tabular}} & \begin{tabular}[c]{@{}l@{}}Agent wants to do \\ the outdoor activity\end{tabular} \\ \hline
\textbf{Emotion} & 1, 2 & \multicolumn{1}{l|}{Very sad} & \multicolumn{1}{l|}{No emotion} & Very happy  \\ \hline
\textbf{\begin{tabular}[c]{@{}l@{}}Emotion \\ trigger 2\end{tabular}} & 
1, 2 & \multicolumn{1}{l|}{Very low trigger} & \multicolumn{1}{l|}{No emotion trigger} & Very high trigger\\ \hline
\textbf{\begin{tabular}[c]{@{}l@{}}Emotion \\ trigger 3\end{tabular}} & 
1, 2 & \multicolumn{1}{l|}{Very low trigger} & \multicolumn{1}{l|}{No emotion trigger} & Very high trigger  \\ \hline
\textbf{Bias} & 1, 2 & \multicolumn{1}{l|}{\begin{tabular}[c]{@{}l@{}}There will be heavy rain\end{tabular}} & \multicolumn{1}{l|}{No bias} & It will be very sunny  \\ \hline
\textbf{\begin{tabular}[c]{@{}l@{}}Rationally \\ perceived \\ knowledge\end{tabular}} & 1, 2 & \multicolumn{1}{l|}{\begin{tabular}[c]{@{}l@{}}There will be  heavy rain\end{tabular}} & \multicolumn{1}{l|}{No information} & It will be very sunny  \\ \hline
\textbf{\begin{tabular}[c]{@{}l@{}}General \\ world \\ knowledge\end{tabular}} & 2 & \multicolumn{1}{l|}{\begin{tabular}[c]{@{}l@{}}Weather prediction \\ is very inaccurate\end{tabular}} & \multicolumn{1}{l|}{\begin{tabular}[c]{@{}l@{}}Weather prediction is \\ mildly accurate\end{tabular}} & \begin{tabular}[c]{@{}l@{}}Weather prediction is\\  very accurate\end{tabular}  \\ \hline
\textbf{\begin{tabular}[c]{@{}l@{}}General\\ preferences\end{tabular}} & 2 & \multicolumn{1}{l|}{\begin{tabular}[c]{@{}l@{}}
Agent strongly dislikes \\ the outdoor activity\end{tabular}} & \multicolumn{1}{l|}{\begin{tabular}[c]{@{}l@{}}
Agent does not have a preference \\ about the outdoor activity\end{tabular}} & \begin{tabular}[c]{@{}l@{}}Agent strongly likes \\ the outdoor activity\end{tabular}  \\ \hline
\textbf{\begin{tabular}[c]{@{}l@{}}Emotion\\ trigger 1\end{tabular}} & 2 & \multicolumn{1}{l|}
{Very low trigger} & \multicolumn{1}{l|}{No emotion trigger} & Very high trigger  \\ \hline
\end{tabular}
\caption{Description of the concepts in real-life scenarios 1 and 2. 
The second column shows in which real-life scenario the concept is present.}
\label{tbl:concepts-real-life-situation}
\end{table}

For implementations, the model core (see \autoref{fig:model_modules}) was considered  
supposing that the rationally perceived knowledge is exactly the same as real-life data. 
Similarly to how humans express their state-of-mind, 
the intensity of beliefs, goals, emotions, and biases were expressed via linguistic terms. 
These terms corresponded to fuzzy values with realisations in $[-1,1]$.   
The links that connect the elements of the network representations 
were also described by linguistic terms  
(according to the expected influence of every pair of connected concepts gathered from the 
intuitive knowledge base from humans) and are quantified according to the rules of fuzzy values.%

The following two real-life scenarios, illustrated in \autoref{FCM:real-life-situations} were considered.  
\emph{Real-life scenario 1}, where the rational agent holds 
a given level of preference for doing an outdoor activity (intensity of the \emph{goal}), 
then checks the weather forecast (\emph{rationally perceived knowledge}),  
and develops a \emph{belief} (that may be biased) about the upcoming weather conditions. 
The agent's belief about the upcoming weather conditions alone, 
or together with the agent's goal (see \emph{emotion trigger 2} and \emph{emotion trigger 3} 
in \autoref{fig:situation1}) may trigger emotions in the agent. 
%
%
\emph{Real-life scenario 2}, where compared to real-life scenario 1, 
 general world knowledge, general preferences, and influence of personality traits are included. 
The rational agent has a \emph{general preference} about the outdoor activity 
and some \emph{general world knowledge} regarding the reliability of the source of the weather forecast. 
When real-life scenario 2 is personalised for human participants, 
the influence of personality traits are also included in the model identification. 
A combination of the agent's general preference and belief (see 
\emph{emotion trigger 1} in \autoref{fig:situation2}) 
may trigger emotions in the rational agent. 
The definitions used for the concepts in real-life scenarios 1 and 2 are given in \autoref{tbl:concepts-real-life-situation}.%

\subsubsection*{Cognitive Model Validation}

The following three hypotheses were assessed and validated for the developed cognitive models:  
\begin{hyp}
\label{hyp:first}
Formulating the weights of the linkages of the  network representations in \autoref{FCM:real-life-situations}  
generally as functions of the concepts 
that correspond to that linkage, rather than considering fixed weights, is essential for accurate estimations 
of the state-of-mind variables of rational agents. 
\end{hyp}
\vspace{-2ex}
\begin{hyp}
\label{hyp:second}
Incorporation of general world knowledge and general preferences   
is essential for accurate estimations 
of the state-of-mind variables of rational agents.  
\end{hyp}
\vspace{-2ex}
\begin{hyp}
\label{hyp:third}
Personalising the weights of the linkages of the network representations 
for each individual (i.e., incorporating the personality traits) is essential for accurate
estimations of the state-of-mind variables of rational agents. 
\end{hyp}

\paragraph{Qualitative assessment based on computer simulations:} 
In order to evaluate Hypothesis~\ref{hyp:first}, weight $w_{12}$ corresponding to the simple linkage between the 
belief, $C_1$, and the goal, $C_2$, (see \autoref{fig:situation1}) is formulated as a function of the belief, i.e.: 
\begin{equation}
    \label{eq:piecewise-function}
    w_{12}(C_1(k)) 
    = \left\{
        \begin{array}{ll}
            w_{12}^- & \quad C_1(k) < 0 \\
            0 & \quad C_1(k) = 0     \\
            w_{12}^+ & \quad C_1(k) > 0
        \end{array}
    \right.
\end{equation}
where $k$ is the simulation step and $w_{12}^- > w_{12}^+$ (for computer-based simulations, we consider 
$w_{12}^- = 0.5 $ and $ w_{12}^+ = 0.1$ 
and for human participants these parameters are personalised per participant). 
More specifically, intuitive human data (also see \cite{Shapiro2007}) revealed that 
compared to a positive belief the influence of a negative belief  
is usually more significant on the (intensity of the) developed goal by rational agents, 
particularly when initially (i.e., before developing a belief) rational agents hold no specific preference regarding their goal. 
For instance, when a person who initially had no specific preference regarding 
doing or not doing an outdoor activity developed the belief that the weather conditions were going to be very bad, 
according to the data collected they were very likely to develop 
the goal of strongly preventing outdoor activities, while after developing the belief that 
the weather conditions were going to be very good 
(although some humans still developed the goal of doing an outdoor activity), 
the goal was much less likely as strong as in the first case. 
This effect was boosted when humans also had a general preference for staying inside.%

\begin{figure}
    \centering
    \begin{subfigure}{0.45\textwidth}
        \centering
        \includegraphics[width=\textwidth]{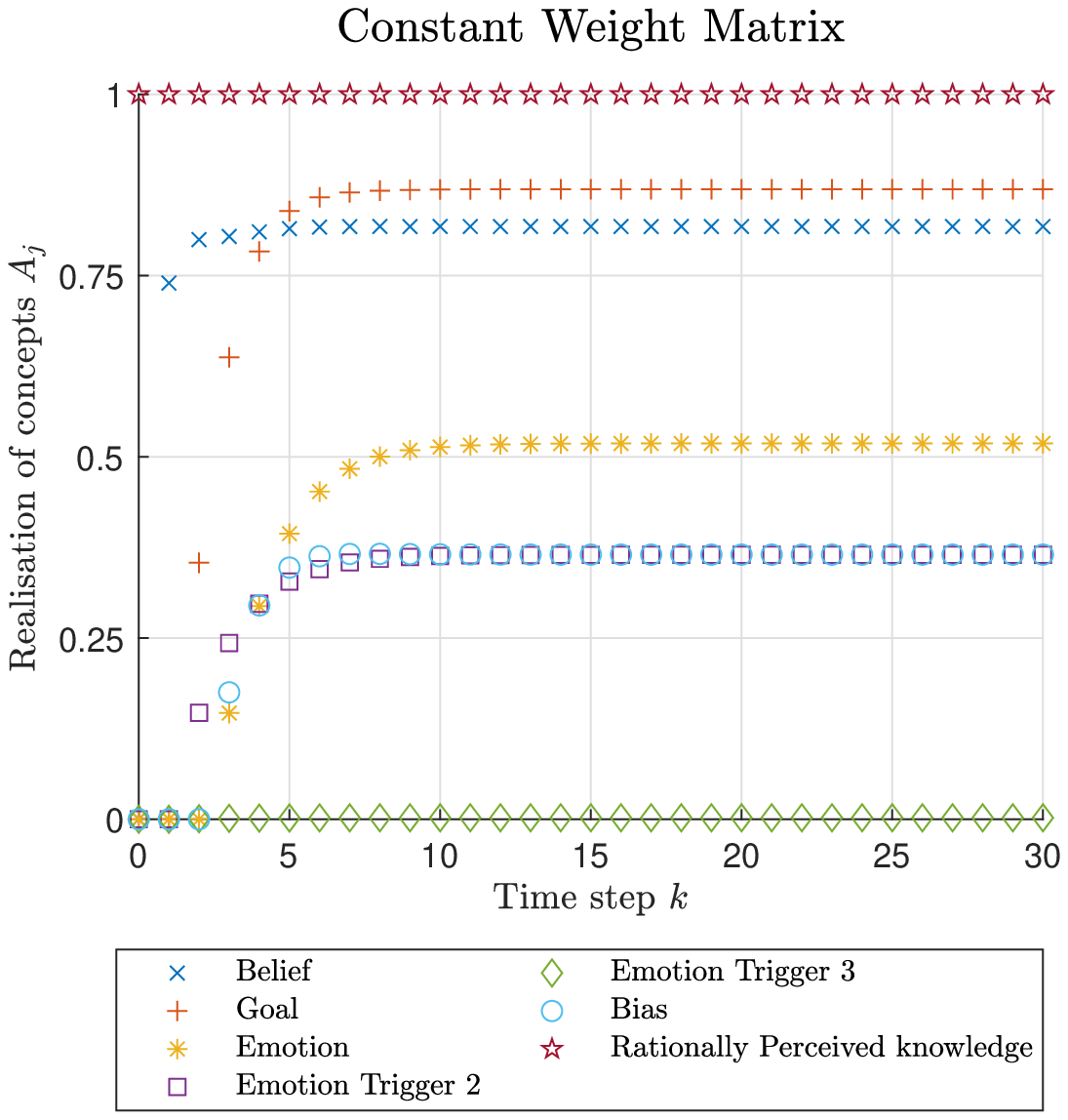}
        \caption{Initial conditions: Null intensity for initial goal ($A_2=0$ for $k=0$) 
        and maximum rationally perceived knowledge ($A_7=1$).}
        \label{fig:h1a-cte-rpk=-1}
    \end{subfigure}   
    \hspace{0.15cm}
    \begin{subfigure}{0.45\textwidth}
        \centering
        \includegraphics[width=\textwidth]{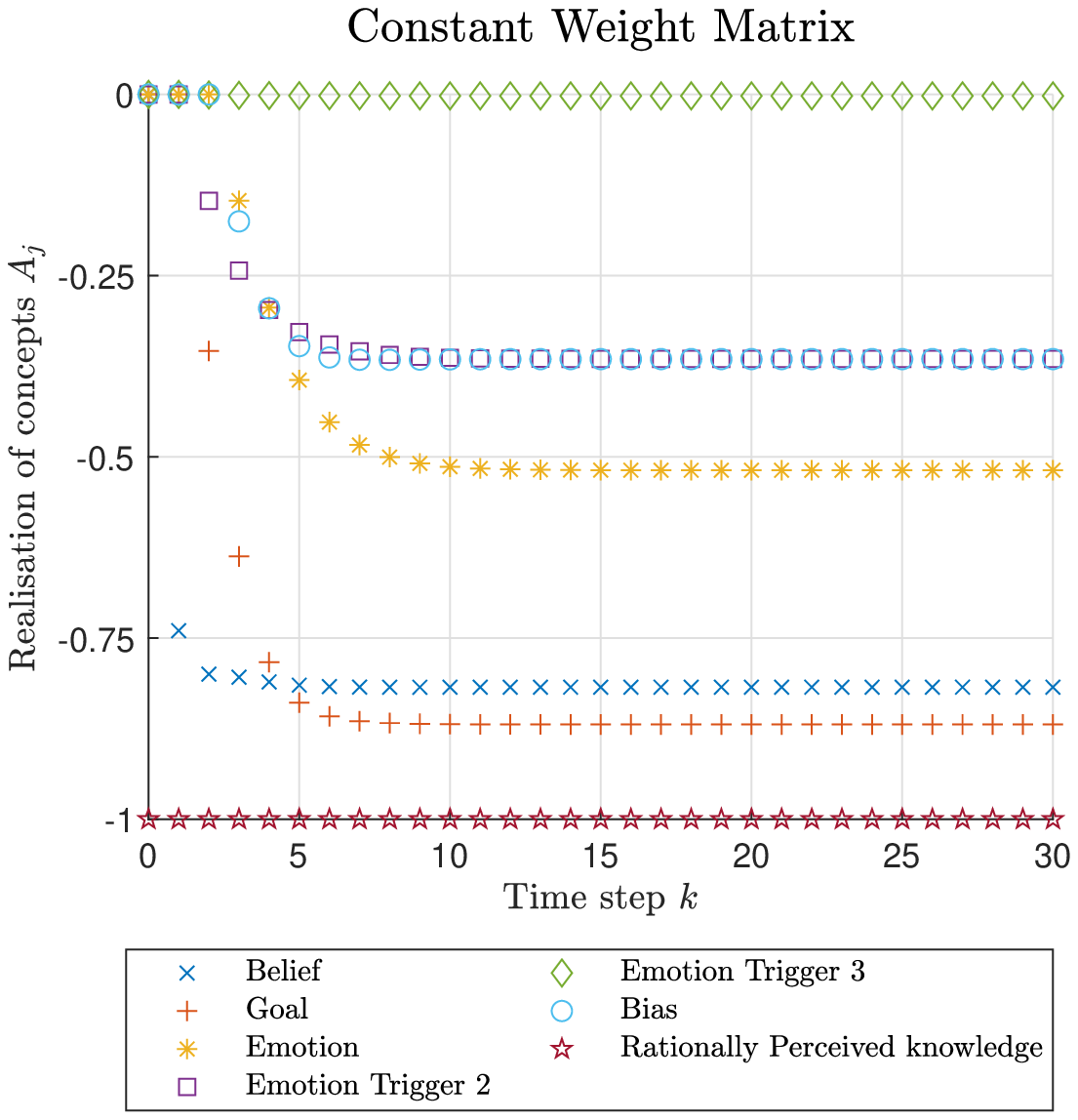}
        \caption{Initial conditions: Null intensity for initial goal ($A_2=0$ for $k=0$) 
        and minimum rationally perceived knowledge ($A_7=-1$).}
        \label{fig:h1a-cte-rpk=1}
    \end{subfigure}
    \centering
    \begin{subfigure}{0.45\textwidth}
        \centering
        \includegraphics[width=\textwidth]{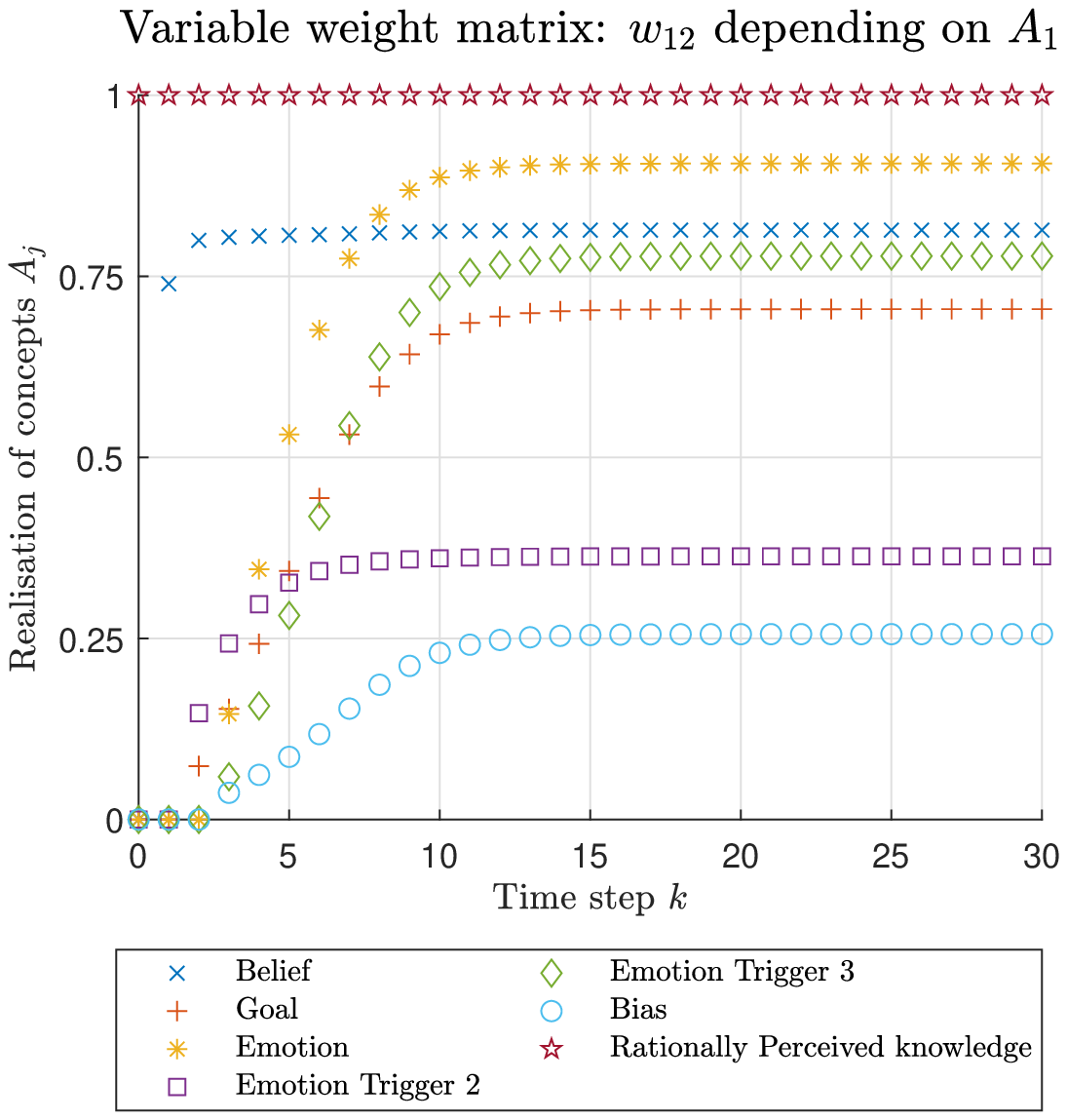}
        \caption{Initial conditions: Null intensity for initial goal ($A_2=0$ for $k=0$) 
        and maximum rationally perceived knowledge ($A_7=1$).}
        \label{fig:h1a-var-rpk=1}
    \end{subfigure}   
    \hspace{0.15cm}
    \begin{subfigure}{0.45\textwidth}
        \centering
        \includegraphics[width=\textwidth]{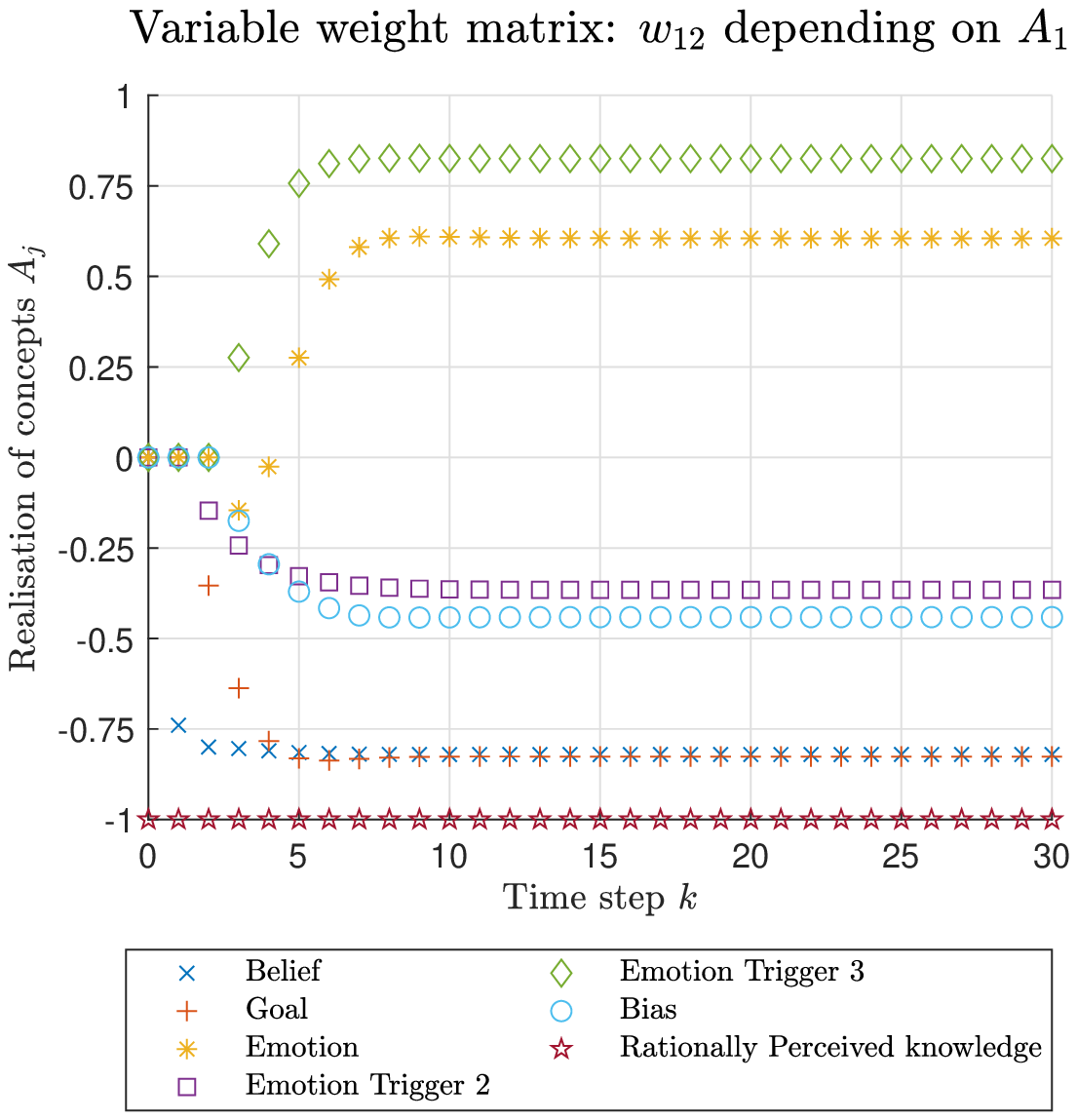}
        \caption{Initial conditions: Null intensity for initial goal ($A_2=0$ for $k=0$) 
        and minimum rationally perceived knowledge ($A_7=-1$).}
        \label{fig:h1a-var-rpk=-1}
    \end{subfigure}
    \caption{Evolution of the model variables over $30$ simulation steps for real-life scenario 1: 
    In the first two cases the weights are constant and in the last two cases weight $w_{12}$  
    varies according to \eqref{eq:piecewise-function}.}
    \label{fig:h1a-var}
\end{figure}

Two sets of simulations for real-life scenarios 1 and 2 
were considered with a null intensity for the initial goal. 
In each simulation set two cases with constant weights for all the linkages and two cases with 
similar initial values, but varying weight $w_{12}$ according to \eqref{eq:piecewise-function} were considered.
For real-life scenario 1 (see \autoref{fig:situation1}) the rationally perceived knowledge   
was once set maximum (i.e., $A_7=1$) and once minimum (i.e., $A_7=-1$), 
implying that the weather conditions were going to be very good 
and very bad, respectively. 
The evolution of the state and auxiliary variables for 
these simulations are shown in \autoref{fig:h1a-var}. 
While for the network representation with constant weights the converged intensity  
of the goal for both very bad and very good weather conditions are the same (see Figures~\ref{fig:h1a-cte-rpk=-1} and \ref{fig:h1a-cte-rpk=1}), for the network representation where $w_{12}$ varies according to \eqref{eq:piecewise-function} 
based on the developed belief, 
the converged intensity of the goal for a negative belief (\autoref{fig:h1a-var-rpk=-1}) 
is larger than a positive belief (\autoref{fig:h1a-var-rpk=1}), 
where these results are in line with reality.%

\begin{figure}
    \centering
    \begin{subfigure}{0.45\textwidth}
        \centering
        \includegraphics[width=\textwidth]{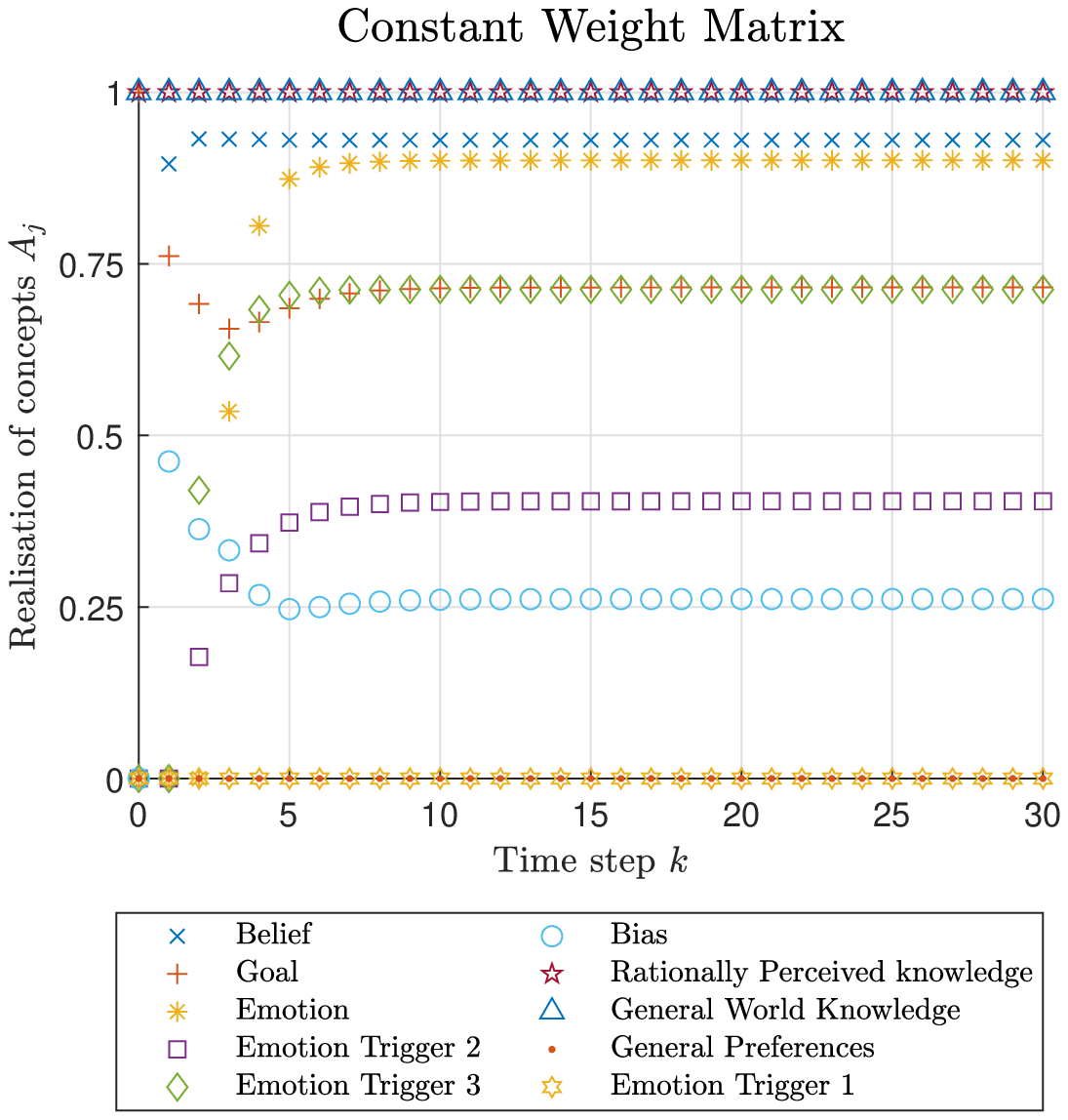}
        \caption{
        Initial conditions: Maximum intensity for initial goal ($A_2=1$ for $k=0$), 
        maximum rationally perceived knowledge ($A_7=1$),  
        maximum general world knowledge ($A_8=1$), 
        and null general preference ($A_9=0$).}
        \label{fig:h1b_cte_gp=0}
    \end{subfigure}   
    \hspace{0.15cm}
    \begin{subfigure}{0.45\textwidth}
        \centering
        \includegraphics[width=\textwidth]{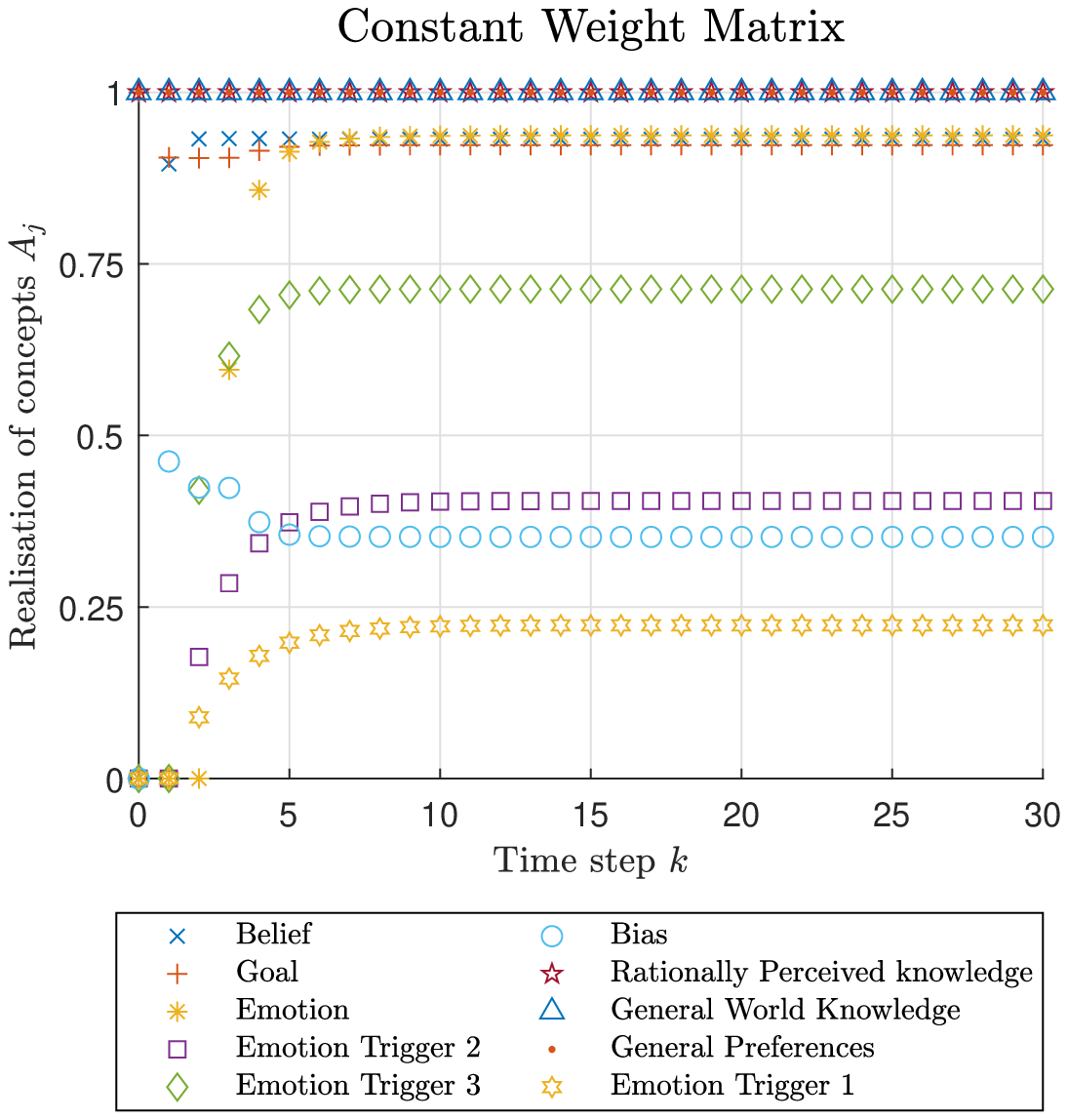}
        \caption{
        Initial conditions: Maximum intensity for initial goal ($A_2=1$ for $k=0$), 
        maximum rationally perceived knowledge ($A_7=1$),  
        maximum general world knowledge ($A_8=1$), 
        and maximum general preference ($A_9=1$)}
        \label{fig:h1b_cte_gp=1}
    \end{subfigure}
%
    \centering
    \begin{subfigure}{0.45\textwidth}
        \centering
        \includegraphics[width=\textwidth]{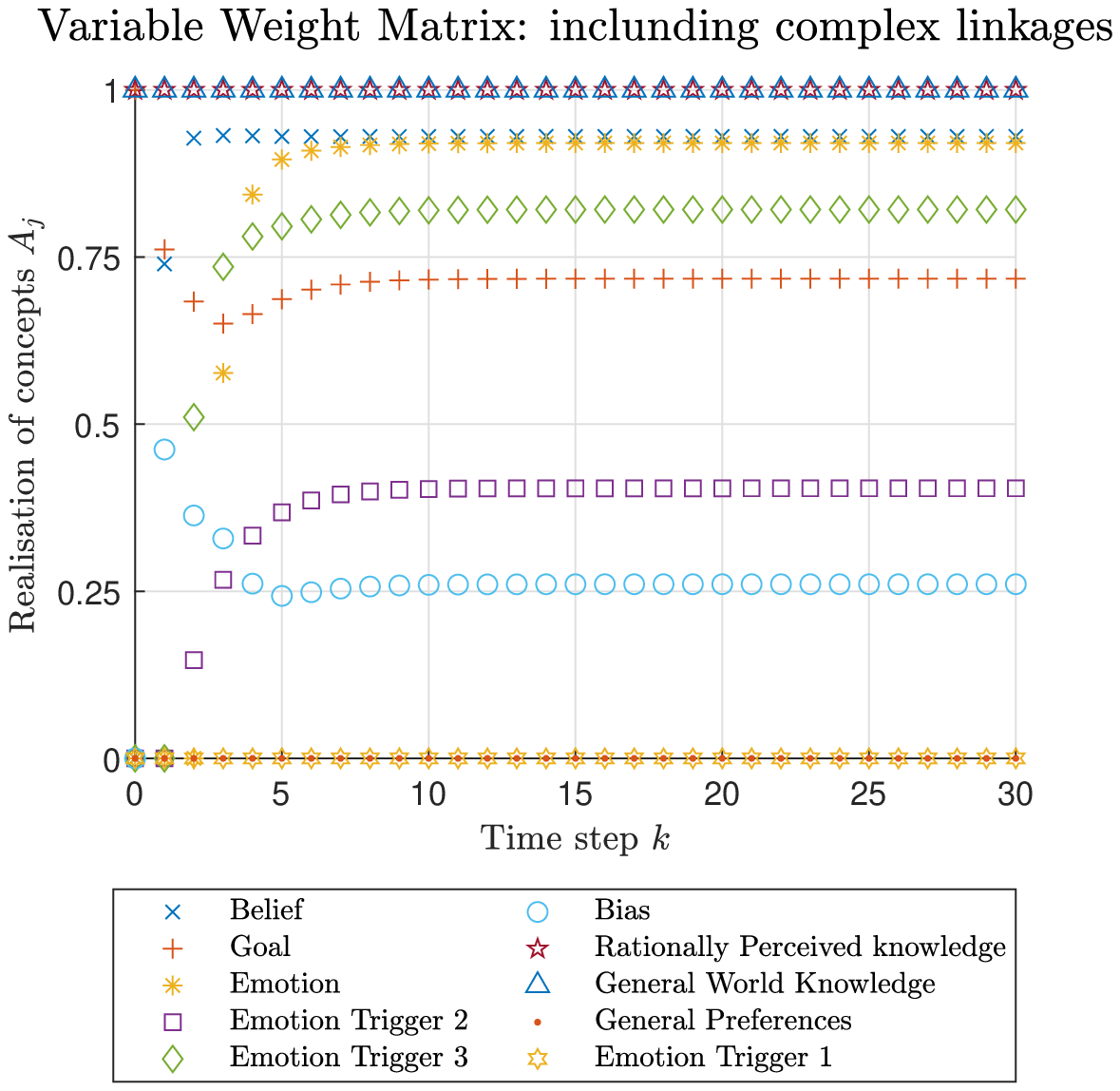}
        \caption{
        Initial conditions: Maximum intensity for initial goal ($A_2=1$ for $k=0$), maximum rationally perceived knowledge ($A_7=1$),  
        maximum general world knowledge ($A_8=1$), and null general preference ($A_9=0$).}
        \label{fig:h1b_var_gp=0}
    \end{subfigure}   
    \hspace{0.15cm}
    \begin{subfigure}{0.45\textwidth}
        \centering
        \includegraphics[width=\textwidth]{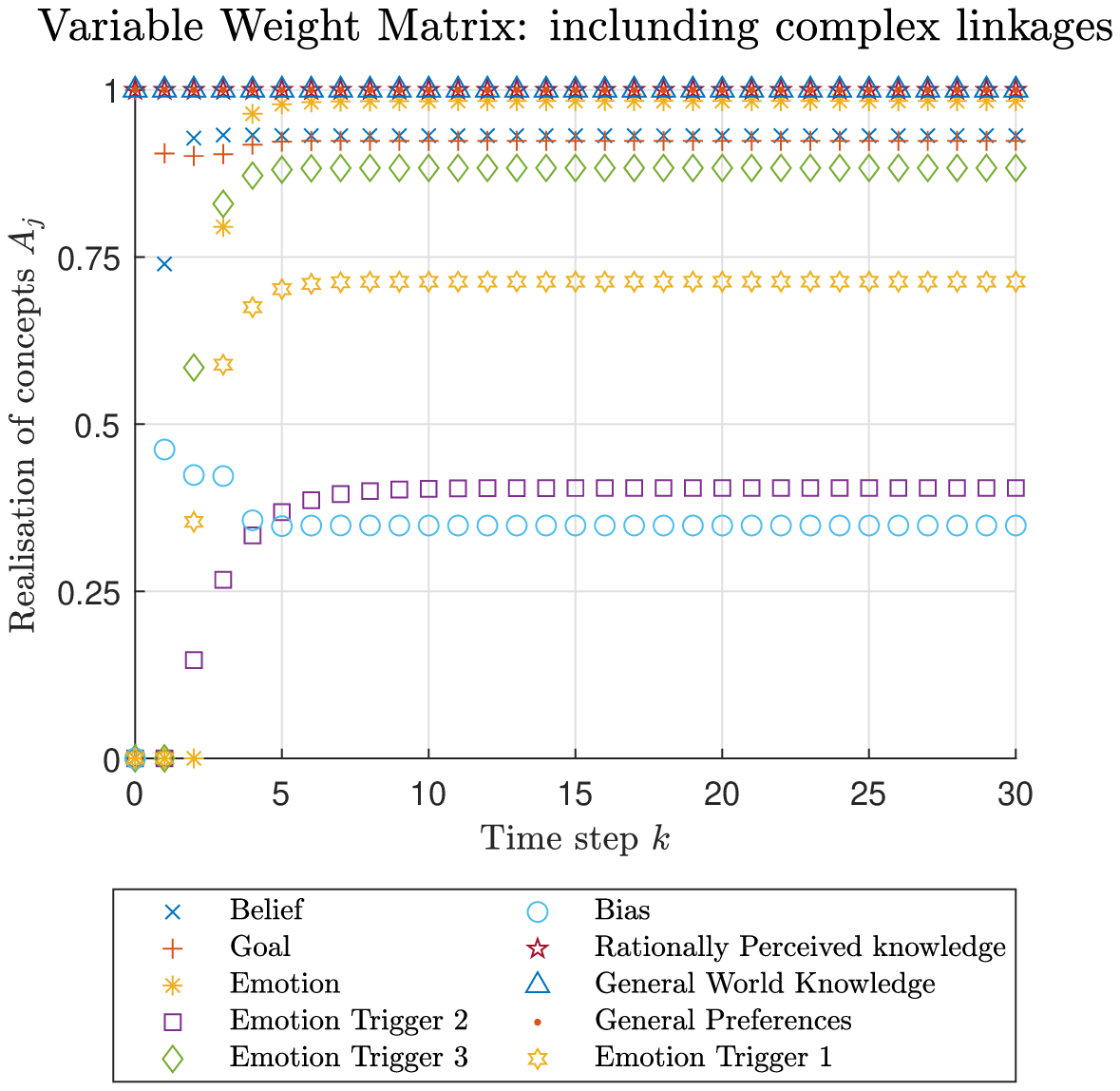}
        \caption{
        Initial conditions: Maximum intensity for initial goal ($A_2=1$ for $k=0$), maximum rationally perceived knowledge ($A_7=1$),  
        maximum general world knowledge ($A_8=1$), and maximum general preference ($A_9=1$).}
        \label{fig:h1b_var_gp=1}
    \end{subfigure}
    \caption{Evolution of the model variables over $30$ simulation steps for real-life scenario 2: 
    In the first two cases, simple linkages (with constant weights) and in the last two cases complex linkages exist.}
    \label{fig:h1b-var}
\end{figure}

\begin{figure}
    \centering
    \begin{subfigure}{0.45\textwidth}
        \centering
        \includegraphics[width=\textwidth]{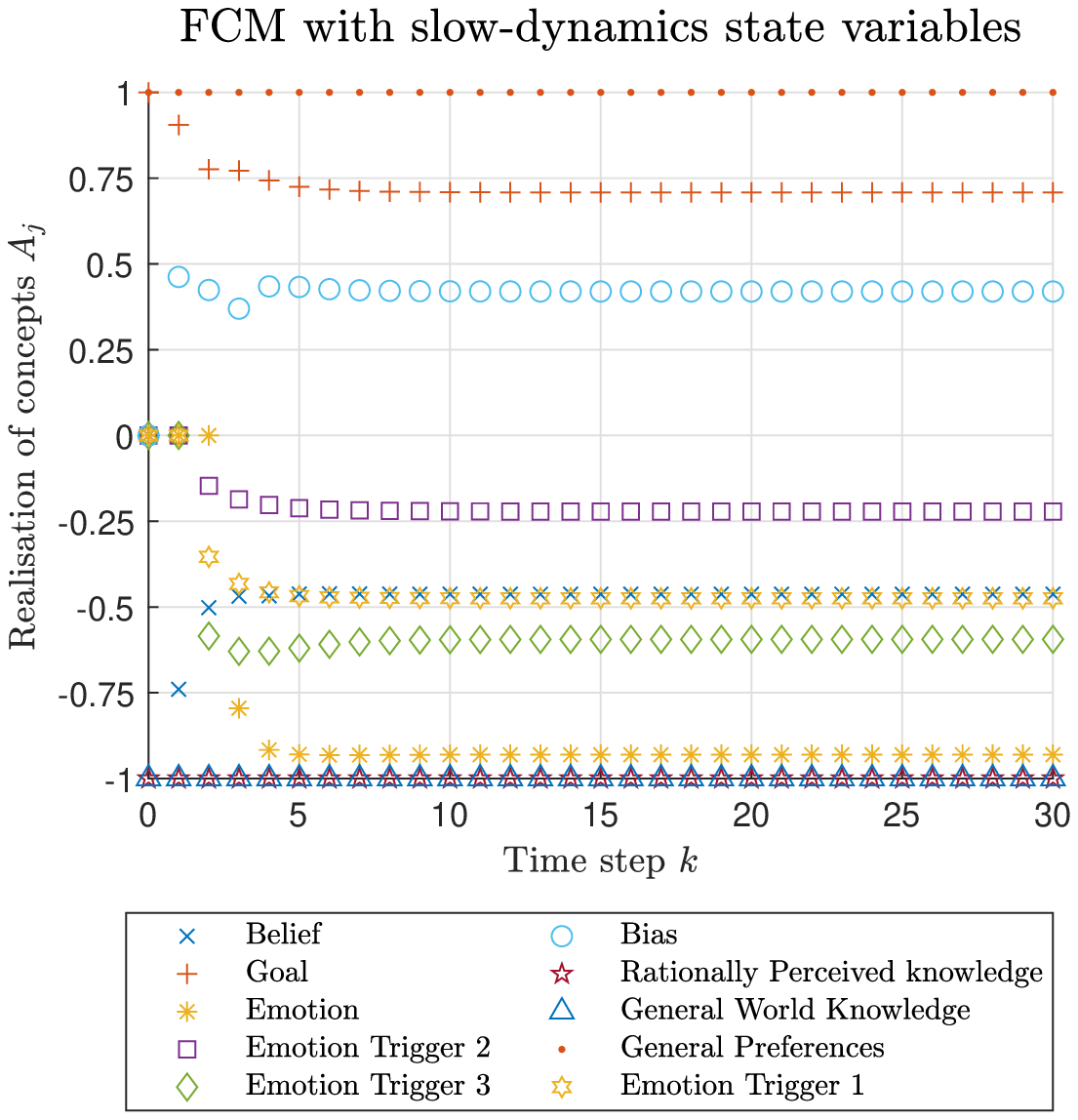}
        \caption{
        Initial conditions: Maximum intensity for initial goal ($A_2=1$ for $k=0$), minimum rationally perceived knowledge ($A_7= -1$),  
        minimum general world knowledge ($A_8=-1$), and maximum general preference ($A_9=1$).}
        \label{fig:h2_gwk-1_gp1}
    \end{subfigure}   
    \hspace{0.15cm}
    \begin{subfigure}{0.45\textwidth}
        \centering
        \includegraphics[width=\textwidth]{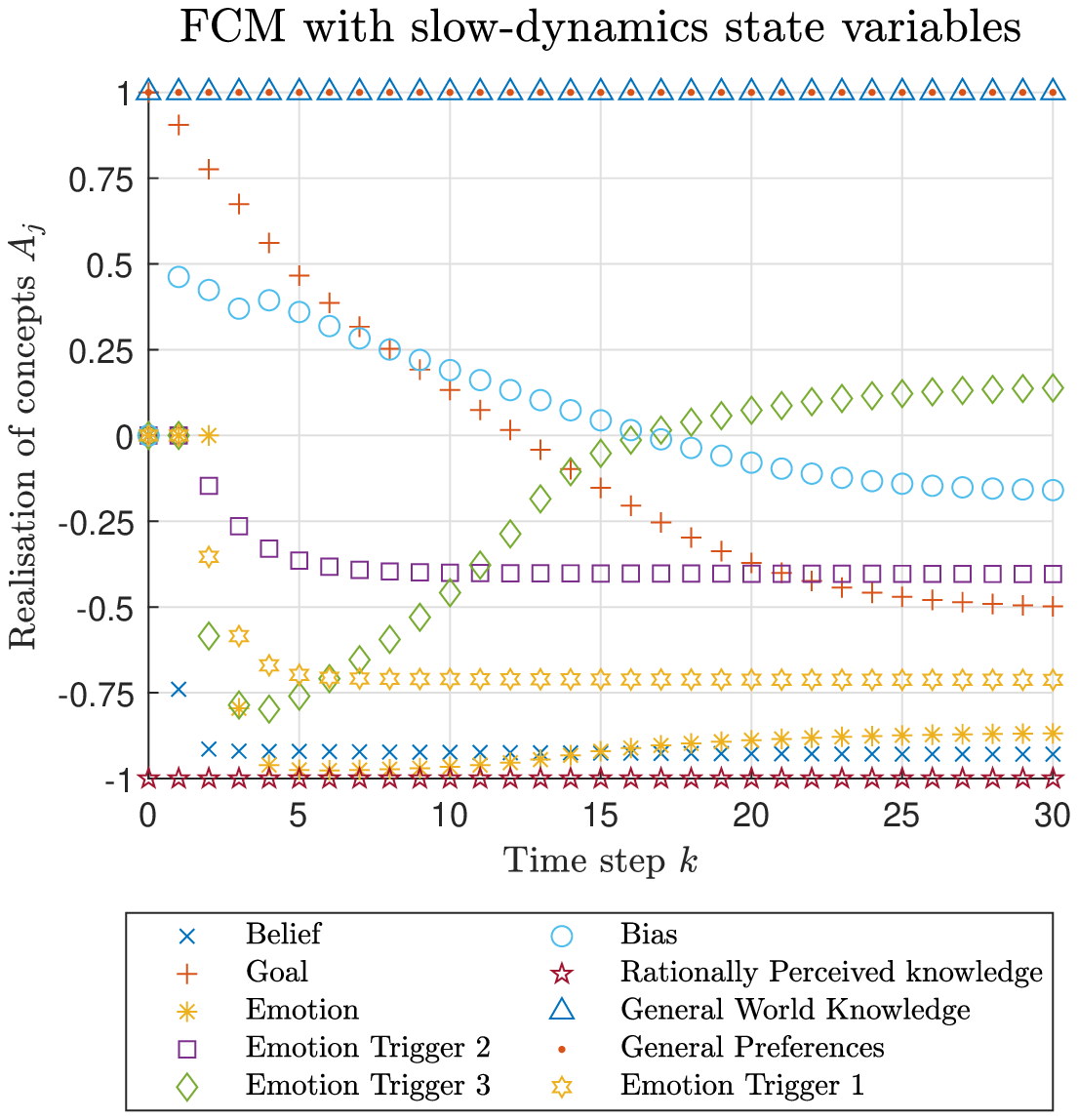}
        \caption{
        Initial conditions: Maximum intensity for initial goal ($A_2=1$ for $k=0$), minimum rationally perceived knowledge ($A_7= -1$),  
        maximum general world knowledge ($A_8=1$), and maximum general preference ($A_9=1$).}
        \label{fig:h2_gwk1_gp1}
    \end{subfigure}
    \begin{subfigure}{0.45\textwidth}
        \centering
        \includegraphics[width=\textwidth]{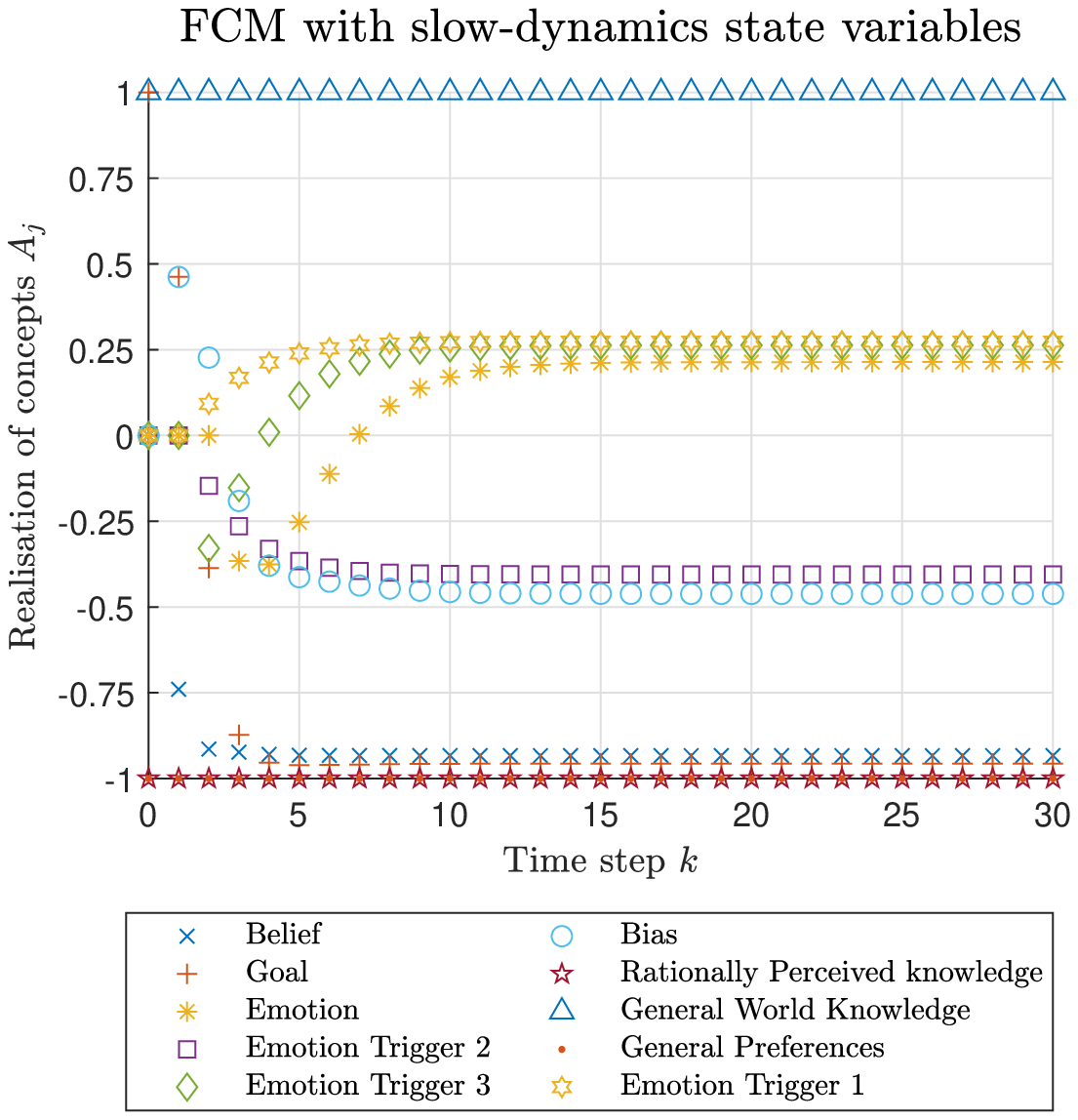}
        \caption{Initial conditions: Maximum intensity for initial goal ($A_2=1$ for $k=0$), minimum rationally perceived knowledge ($A_7= -1$),  
        maximum general world knowledge ($A_8=1$), and minimum general preference ($A_9=-1$).}
        \label{fig:h2_gwk1_gp-1}
    \end{subfigure}
    \caption{Evolution of the model variables over $30$ simulation steps for real-life scenario 2}
    \label{fig:h2-var}
\end{figure}

The next case that was simulated was based on the following observation: 
When humans strongly wanted to do an outdoor activity, and 
developed the belief that the weather conditions were going to be very good, 
emotion trigger 3 (see Figures~\ref{fig:situation1} and \ref{fig:situation2}) 
had a higher intensity than when the weather conditions were the same, 
but such a strong intensity regarding doing an outdoor activity did not exist (also see \cite{Harris1989,Bradmetz2004,Reisenzein2009}). 
In other words, while beliefs, either alone or together with goals, may trigger emotions, 
their influence over emotions increases (decreases) with the intensity of the goal increasing (decreasing). 
%
This is represented via a complex linkage in the proposed network representation (see \autoref{FCM:real-life-situations}), 
i.e., goal is an intermediate concept that influences the weight corresponding to the direct linkage 
between the belief and emotion trigger 3. 
In other words, this linkage should mathematically be represented as a function of (at least) 
the intermediate concept, goal. 
When general world knowledge and particularly general preferences were excluded 
(see \autoref{fig:situation1}), the variables of the cognitive model converged rapidly to similar values 
in both cases of the given example. 
Therefore, the network representation shown in \autoref{fig:situation2} was used 
to simulate the discussed example, where  
the weather conditions (i.e., rationally perceived knowledge) were going to be very good (i.e., $A_7=1$), 
according to the general world knowledge the accuracy of the weather forecast was very high (i.e., $A_8=1$), 
and the rational agent initially held the highest intensity for doing an outdoor activity (i.e., $A_2=1$ for $k=0$). 
Once the general preference was set to $A_9=0$ and once to $A_9=1$. 
%
The simulation results for constant and varying weights for the linkages are shown in \autoref{fig:h2-var}:   
In Figures~\ref{fig:h1b_cte_gp=0} and \ref{fig:h1b_cte_gp=1} (constant weights),   
there is yet no significant difference in the estimated emotion of the rational agent for the two different cases. 
With varying weights however (see Figures~\ref{fig:h1b_var_gp=0} and \ref{fig:h1b_var_gp=1}), 
the converged value for emotion trigger 3 and thus emotions in the first case  
 is lower than for  the  second case, providing a more realistic simulation. 
 These findings support both Hypothesis~\ref{hyp:first} and Hypothesis~\ref{hyp:second}.%

%


Next we considered two cases where despite similar rationally perceived knowledge and general preferences, 
the general world knowledge differs.  
We supposed very bad weather conditions (i.e., $A_7=-1$), and the rational agent 
initially holding the goal of doing an outdoor activity with the maximum intensity (i.e., $A_2=1$ for $k=0$) 
and a strong general preference regarding outdoor activities (i.e., $A_9=1$). 
Once the agent considered the weather forecast to be very accurate (i.e., $A_8=1$)  
and once to be very inaccurate (i.e., $A_8= -1$). 
The simulation results using the network representation shown in \autoref{fig:situation2}  
are presented in \autoref{fig:h2-var}: 
The cognitive model captured the difference in the evolution of the belief 
of the rational agent, i.e., whenever the rational agent considered the source of the weather forecast  
to be unreliable, the converged value of the belief was lower than the rationally perceived knowledge,  
while for a very reliable weather forecast, the belief's value converged to the rationally perceived knowledge (see \autoref{fig:h2_gwk1_gp1}). 
This difference cannot be captured via the model shown in \autoref{fig:situation1}.  
In addition to the belief, emotions are often influenced by 
the general preference and general world knowledge: 
\autoref{fig:h2_gwk1_gp-1} illustrates that   
a rational agent who initially held the goal of doing an outdoor activity with maximum intensity 
but minimum general preference, and realised via a highly reliable weather forecast source that 
it was going to be very bad weather conditions, 
according to the network representation in \autoref{fig:situation2} 
ended up with a small positive value for the emotion (e.g., slightly happy). 
Moreover, the intensity of the goal converges to approximately $-0.5$, implying that 
the rational agent keeps a medium intention 
for going outside despite the very bad weather conditions. 
These realistic results, which cannot be obtained via the network representation in \autoref{fig:situation1}, 
further support the validity of Hypothesis~\ref{hyp:second}. 
In order to evaluate the validity of Hypothesis~\ref{hyp:third},  
detailed data from $15$ individual participants were used to personalise the cognitive models. 
The details are given next.%

\paragraph{Quantitative assessment based on human participants:}
The validity of Hypotheses~\ref{hyp:first}-\ref{hyp:third} was further investigated 
via the results of an online survey that compared the beliefs, goals, and emotions of human participants 
in various real-life scenarios with those computed by the proposed cognitive models. 
Overall, $15$ participants took part in the survey, 
which considered $26$ scenarios. 
The general preferences, rationally perceived knowledge, 
and general world knowledge    
were initialised as inputs of the cognitive models and survey.  
\autoref{tbl:scenario-inputs} shows the linguistic terms (provided by participants) for the inputs in the online survey 
and their corresponding values use by the cognitive models.  
The following four cognitive models were considered: 
\begin{compactitem}
    \item 
    \textbf{Model 1}, which included the model core shown in \autoref{fig:situation2}, 
    was personalised per participant, and the linkages where represented by functions. 
    \item
    \textbf{Model 2}, which was based on \autoref{fig:situation1} (excluding 
general preferences and general world knowledge), was personalised per participant, 
and the linkages were represented by functions. 
    \item 
    \textbf{Model 3}, which included the model core shown in \autoref{fig:situation2},  
    was personalised per participant, 
    and the linkages were represented by constant weights.  
    \item
    \textbf{Model 4}, which included the model core shown in \autoref{fig:situation2}, 
    was \emph{not} personalised for various participants, 
    and the linkages were represented by functions. 
\end{compactitem}
The $26$ scenarios were divided into $6$ sets, where in each only $1$ out of the $3$ inputs varied 
(i.e., rationally perceived knowledge for sets $1$ and $2$, 
general world knowledge for sets $3$ and $4$, 
general preferences for sets $5$ and $6$). 
Sets $1$-$4$ each included $3$ scenarios, while sets $5$ and $6$ included $7$ scenarios. 
The format of the questions asked in the online survey is briefly explained in \autoref{app:survey}. 
Around $70\%$ of the data collected via the online survey was used for training 
and the rest $30\%$ for validation. 
Since the number of data was limited, 
each model was trained and validated using three different batches of training and validation scenarios. 
First the cognitive models were trained via sets $3$-$6$ and were validated via sets $1$ and $2$. 
Second, sets $1$, $2$, $5$, and $6$ were used for training and sets $3$ and $4$ for validation. 
Lastly, sets $1$-$5$ and set $6$ were respectively used for training and for validation. 
For model $m$ with $m=1,2,3$ the weights/functions of those linkages that are regulated by personality traits 
(i.e., linkages influencing the emotion triggers and the bias and the goal via the belief) 
were identified per participant via a grid search that determined vector 
$\bm{w}$ of constant weights or function parameters that for participant $p$ minimised the following loss function:
\begin{equation}
    \label{eq:loss-function}
    J_{m,p}(\bm{w}) = 
    \sum_{s\in\mathbb{S}_{\textrm{t}}}
    \bigg(
    \sum_{j= 1,2,3}
    \Big( 
    A_{m,j}(\bm{w},s) - A_{p,j}(s)
    \Big) ^2 
    \bigg)
\end{equation}
with $\mathbb{S}_{\textrm{t}}$ the set of scenarios used for training, 
$A_{m,j}(\bm{w},s)$ the converged value of the realisation of concept $C_j$ estimated via model $m$  
using weight vector $\bm{w}$ for scenario $s$, 
and $A_{p,j}(s)$ the quantified value within $[-1,1]$ corresponding to the linguistic response 
 of participant $p$ for concept $C_j$ in scenario $s$. %
The converged values of the state variables (belief, goal, emotion) estimated by model $m$   
for scenario $s$ was compared with the corresponding values from the online survey.  
The mean squared error for realisation $A_j$ of concept $C_j$ for 
scenario $s$ for model $m$ and for $j=1,2,3$ is:
\begin{align}
    \label{eq:evalution-metrics-scenario}
    \text{MSE}(m,j,s) = \dfrac{1}{|\mathbb{P}|}
    \sum_{p\in\mathbb{P}}
    \bigg(A_{m,j}(\bm{w}^*_{m,p}, s)-A_{p,j}(s)\bigg)^2 
\end{align}
with $\bm{w}^*_{m,p}$ the vector of personalised weights for model $m$ obtained via \eqref{eq:loss-function} 
for participant $p$  and  $|\cdot|$ the set cardinality. 
The mean squared error of model $m$ for concept $C_j$ and validation set $\mathbb{V}$ is:
\begin{align}
    \label{eq:evalution-metrics-overall}
    \text{MSE}\left(m,j,\mathbb{V}\right) 
    = \dfrac{1}{|\mathbb{V}|} 
    \sum_{s\in\mathbb{V}}  \text{MSE}(m,j,s)
\end{align} 
The same procedure was used to identify model $4$ but the loss function in \eqref{eq:loss-function} 
was considered for the entire population of participants at once.%


\begin{table}
\centering
\begin{tabular}{|l|c|c|}
\hline
\textbf{Concept} & \textbf{Linguistic term} & \textbf{\begin{tabular}[c]{@{}c@{}}Numerical realisation \end{tabular}} \\ \hline
\multirow{7}{*}{\textbf{\begin{tabular}[c]{@{}l@{}}General preferences\end{tabular}}} & Dislike a great deal & -1 \\ \cline{2-3} 
  & Dislike a moderate amount & -0.66\\ \cline{2-3} 
  & Dislike a little & -0.33\\ \cline{2-3} 
 & No preference & 0 \\ \cline{2-3} 
  & Like a little & 0.33\\ \cline{2-3} 
 & Like a moderate amount & 0.66 \\ \cline{2-3} 
 & Like a great deal & 1 \\ \hline
\multirow{5}{*}{\textbf{\begin{tabular}[c]{@{}l@{}}Rationally perceived knowledge\end{tabular}}}  & Heavy rain & -1\\ \cline{2-3} 
  & Light rain & -0.5\\ \cline{2-3} 
  & Unknown & 0\\ \cline{2-3} 
  & Cloudy & 0.5\\ \cline{2-3} 
  & Sunny & 1\\ \hline
\multirow{3}{*}{\textbf{\begin{tabular}[c]{@{}l@{}}General world knowledge\end{tabular}}}  & Inaccurate & -0.4\\ \cline{2-3} 
  & Accurate & 0.2\\ \cline{2-3} 
  & Very accurate & 0.8\\ \hline
\end{tabular}
\caption{Linguistic terms and their values considered as the inputs (general preferences, 
rationally perceived knowledge, and general world knowledge) in the online survey.}
\label{tbl:scenario-inputs}
\end{table}


\begin{table}
\centering
\begin{tabular}{|c|c|c|c|c|}
\hline
\textbf{Belief} & \textbf{First run} & \textbf{Second run} & \textbf{Third run} & \textbf{All the validation data} \\ \hline
\textbf{Model 1} & 0.473 & 0.148 & 0.057 & 0.218 \\ \hline
\textbf{Model 2} & 0.541 & 0.234 & 0.059 & 0.264 \\ \hline
\textbf{Model 3} & 0.493 & 0.393 & 0.085 & 0.314 \\ \hline
\textbf{Model 4} & 0.549 & 0.156 & 0.105 & 0.261 \\ \hline

\hline
\textbf{Goal} & \textbf{First run} & \textbf{Second run} & \textbf{Third run} & \textbf{All the validation data} \\ \hline
\textbf{Model 1} & 0.386 & 0.400 & 0.115 & 0.314 \\ \hline
\textbf{Model 2} & 0.426 & 0.793 & 0.133 & 0.461 \\ \hline
\textbf{Model 3} & 0.548 & 0.744 & 0.418 & 0.593 \\ \hline
\textbf{Model 4} & 0.373 & 0.406 & 0.219 & 0.353 \\ \hline

\hline
\textbf{Emotion} & \textbf{First run} & \textbf{Second run} & \textbf{Third run} & \textbf{All the validation data} \\ \hline
\textbf{Model 1} & 0.185 & 0.198 & 0.120 & 0.162 \\ \hline
\textbf{Model 2} & 0.255 & 0.268 & 0.183 & 0.234 \\ \hline
\textbf{Model 3} & 0.244 & 0.288 & 0.141 & 0.223 \\ \hline
\textbf{Model 4} & 0.211 & 0.257 & 0.128 & 0.195 \\ \hline
\end{tabular}
\caption{Mean squared error in estimation of the belief, goal, and emotion via models 1-4 for 
three different validation runs. 
The last column shows the average error for all validation scenarios.}
\label{tbl:results-emotions}
\end{table}

Table~\ref{tbl:results-emotions} shows the mean squared error for estimation of 
the beliefs, goals, and emotions via models $1$-$4$. 
Since all realisations were already normalised within $[-1,1]$, absolute error values were used 
and the mean squared error was maximum $4$.   
Compared to model $3$, model $1$ had significantly lower errors 
for all validation sets for the three state variables, which supports the validity of Hypothesis~\ref{hyp:first}, 
i.e., the importance of formulating the linkages of the model as functions. 
Compared to model $2$, model $1$ had lower errors for all validation sets and 
and state variables, which supports the validity of Hypothesis~\ref{hyp:second}, i.e., 
the importance of including general preferences and general world knowledge in the cognitive models. 
Finally, compared to model $4$, model $1$ made more accurate estimates of beliefs and emotions for all validation sets. 
The error in estimation of the goal was close for both models for the first two validation sets, and 
was lower for model $1$ for validation set $3$. 
These results support the validity of Hypothesis~\ref{hyp:third}, i.e., the importance of personalising the cognitive models. 
The satisfactory average performance of model $4$ (see the last columns of Table~\ref{tbl:results-emotions}), 
implies that such a universal cognitive model 
can reliably be used when personal information is not yet available.%

\section{Conclusions and Topics for Future Research}
\label{sec:conclusions}

We formalised various cognitive procedures of humans via network representations, proposed an extended version of fuzzy cognitive maps, and formulated the corresponding mathematical cognitive models for these network representations. 
While previous research focuses on evolution of beliefs and goals only, we considered emotions and personality traits, which allow for incorporation of biases in perception and beliefs. 
We performed several analyses based on realistic experiments that included cognitive procedures of humans 
in order to personalise and evaluate the accuracy and validity of the proposed cognitive models. 
Beliefs, goals, and emotions were considered as the model's state variables that evolve in the course of small time scales (e.g., seconds or minutes).  
General world knowledge, general preferences, and personality traits 
that evolve in much larger time scales (e.g., months or years)  
were included as constant parameters within the small time scales of the modelled dynamics. 
While previous research (see \cite{Kwon2008, Saxe2017,Tapus2008,Zaki2013, Lee2019,Bono2007PersonalitySelf-monitoring,Leite2013}) 
shows the importance of emotions and personality traits in ToM, this is the first time these elements are systematically included in mathematical cognitive models for humans.  
The resulting cognitive models were identified and validated based on computer-based 
simulations and real-life experiments with human participants. 
The results of these experiments showed that the proposed cognitive models successfully 
represent personal differences of participants and precisely estimate and predict 
their current and future state-of-mind and behaviors. 
Moreover, the results proved that including the emotions and personality traits, and thus incorporating personalisation and biases that exist in real-life cognitive procedures of humans, 
as well as including the general world knowledge 
and general preferences are essential for realistic, precise, and personalised estimation 
of the unobservable state-of-mind of humans.%

In the future, the proposed cognitive models will be used as prediction models of humans for control systems that 
steer the behaviour of autonomous machines that interact with humans.  
Additionally, such models, when personalised to represent the cognitive procedures of an expert 
(e.g., a therapist), may be used to exhibit expert-like interactive behaviours via autonomous machines.%

\section*{Compliance with Ethical Standards}

\smallskip
\noindent
\textbf{Competing Interests:} 
The authors declare no competing interests.

\bigskip
\noindent
\textbf{Research involving Human Participants and/or Animals:} 
The research involved volunteer human participants. 
The research has been approved by the Human Research Ethics Committee of TU Delft.

\bigskip
\noindent
\textbf{Informed Consent:} 
The participants were completely informed about and agreed with the data that was collected via the surveys and that this data will remain anonymous. 

\bigskip
\noindent
\textbf{Author Contributions:} 
Author M.\ Mor\~{a}o Patr\'{i}cio  
contributed to designing and implementing the computer-based and real-life experiments. 
Authors M.\ Mor\~{a}o Patr\'{i}cio and A.\ Jamshidnejad 
contributed to the analysis and interpretation of the results, 
development of mathematical cognitive models of humans, 
and composition of the manuscript. 
Author A.\ Jamshidnejad supervised the study
design and edited the manuscript. 
Both authors have critically reviewed the manuscript
and approved the final version of the manuscript.

\bibliographystyle{spbasic_custom}      
\bibliography{Reference_Ana.bib}   


\newpage
\appendix

\section{Examples}
\label{app:examples}

This appendix represents a number of examples according to real-life scenarios, 
where these examples motivate the elements and/or linkages that are introduced in the proposed cognitive model.%
\medskip

\noindent
\textbf{Role of slow-dynamics state variables:} 
The following three examples show the influence of the slow-dynamics state variables on the dynamic evolution of fast-dynamics state variables. In these examples, the observer agent (described by first-person pronouns) makes inferences about the 
state-of-mind of the observed agent (referred to by a specific name) based on their observed actions (\emph{inverse inference}).%

\begin{example}
\label{example:belief_guess}
Ana and I are both in the library at 5:30 PM. 
Ana picks up her wallet and walks towards the door (\textit{action of the observed agent noticed by the observer agent}). 
The coffee house nearby has a late opening hour until 6:00 PM, but I do not know whether or not Ana knows about this 
(no access to the general world knowledge of the observed agent).
I guess Ana knows about the opening hour of the coffee house and believes that it is still open (\textit{guessing the general world knowledge and inferring about the belief of the observed agent}). 
I infer that Ana is going to buy a cup of coffee (\textit{goal of the observed agent inferred by the observer agent}).%
\end{example}
In this example, since the observer agent does not have access to the general world knowledge of the observed agent,  
the inference involves an  intermediate procedure, i.e., inference about the belief of the observed agent  
based on a guess rather than facts.%

Suppose that according to Ana's general world knowledge the opening hour of the nearby coffee house is until 5:00 PM. 
Thus Ana does not go at 5:30 PM to the coffee house, which she supposes to be closed. 
In case I had access to Ana's general world knowledge, then 
I would not conclude that Ana is going to grab a coffee. 
Therefore, my inference about Ana's goal was more reliable.%

\begin{example} 
\label{example:belief_inference}
Consider the scenario of Example~\ref{example:belief_guess}, but this time I have heard before from Ana 
that she knows the coffee house nearby is open until 6:00 PM 
(\textit{access to the general world knowledge of the observed agent}). 
This time I infer with more certainty that Ana is going to buy a cup of coffee.%
\end{example}
Although this example shows that when the general world knowledge of the observed agent is known by 
the observer agent, an inverse inference about the goals of the observed agent are less prone to uncertainties,  
the next example shows that knowing the general world knowledge of the observed agent alone may not 
suffice to make precise inverse inferences.%

\begin{example}
\label{example:belief_GWK}
Suppose that in Example~\ref{example:belief_inference}, in addition to being aware of Ana's general world knowledge, 
I know she needs to drink coffee when studying late (\textit{access to the general preference of the observed agent}).  
Then I infer that Ana is going to buy coffee with a much higher certainty than in Example~\ref{example:belief_inference}.   
On the contrary, if I know that Ana never drinks coffee in the afternoon (\textit{general preference of the observed agent})  
I will not infer that her beliefs and goals are related to grabbing a coffee. 
\end{example} 
This example shows that the access of an observer agent to the general world knowledge and general preferences 
of an observed agent significantly improves the reliability and level of certainty of the inferred fast-dynamics state variables. 
Moreover, having access to only one of these slow-dynamics state variables may still result in inaccurate or erroneous inferences. 
In particular, having access to the general preferences of the observed agent in addition to the general world knowledge 
in \autoref{example:belief_GWK} supports the certainty of the inferences or prevents the observer agent from making 
erroneous inferences.%

\begin{example}
\label{example:belief_personalitytraits}
Now consider \autoref{example:belief_GWK}, where I am aware that Ana never drinks coffee in the afternoon. 
While the combination of Ana's general world knowledge and general preference prevents me from making a wrong inference, 
they do not provide me with a chance either to make an inference about what Ana's beliefs and goals at the moment are. 
Now suppose that I know Ana for long enough to be aware that she is an introvert (i.e., she needs some personal time 
after long interactions) with higher levels of neuroticism\footnote{We use the terms introversion and neuroticism 
according to the categorisation introduced by the Big Five Personality Traits \citep{BigFivePersonalities}.} 
(i.e., she often feels worried). These personality traits of Ana together with her ctions   
make me to infer that Ana believes she needs some alone time (thus leaving me for a few minutes to be by herself). 
I also infer that Ana believes if she leaves her wallet unattended, someone may steal her credit or ID cards  
(thus taking her wallet with her).
\end{example}
This example shows the importance of incorporating the specific personality traits of rational agents 
in cognitive models for achieving more precision and reliability with the resulting inferences.%
\medskip

\noindent
\textbf{Elements that influence emotions:} 
The following three examples illustrate the role of various state variables in triggering the emotions of an observed agent 
(referred to by a specific name).%

\begin{example}
\label{example:belief_generating_emotions}
While walking on the street, Elisa's wallet falls out of her purse (\textit{real-life data}). Later on in a shop 
Elisa reaches for her wallet and realises that it is not in her purse (\textit{perceptual access}). 
She reasons that she has lost the wallet (\textit{rationally perceived knowledge}). She then supposes that she has lost her wallet (\textit{inference of a belief based on the rationally perceived knowledge}). 
This belief makes her anxious (\textit{stimulation of emotions}). 
\end{example}
In the given example, before Elisa notices that her wallet is missing (i.e., \textit{without perceptual access}) and reasons that she has lost it (i.e., \textit{without rational reasoning}), she was not anxious (\textit{no stimulation of emotions}). 
In a different situation, for the same perceptual access that causes the same perceived data, i.e., a missing wallet, Elisa may reason and believe that she has left her wallet on the dining table at home (\textit{different rational reasoning and hence different rationally perceived knowledge}). 
Therefore, Elisa will not be anxious (\textit{no stimulation of emotions}). 
In summary, independent of what the real-life data is (e.g., the wallet has fallen on the street or is at home)  the emotions of a rational agent may be  moderated by the perceptual access of the agent to that data and by the reasoning it applies to the perceived data. 
In other words, the emotions of a rational agent depend on its beliefs rather than on real-life data directly.%


\begin{example}
\label{example:goal_to_emotion}
Frank is exploring a new city for the first time and wants to buy an ice cream (\textit{goal}). 
While walking, he notices a few people 
across the street who are eating ice cream (\textit{perceived data}). 
Correspondingly, he reasons and believes that there should be an ice cream shop close by (\textit{rationally perceived knowledge transformed into a belief}), 
which makes him feel satisfied (\textit{stimulated emotions}).%
\end{example}
This example shows a case where a belief by itself does not stimulate emotions, but the belief together with a 
goal does. In other words, if Frank did not want to eat an ice cream, the belief that an ice cream shop is nearby 
would not influence his emotional status.%
%

\begin{example}
\label{example:general_preference_to_emotions}
 Grace is afraid of dogs (\textit{general preference}). 
 While walking in a park, she notices the footprints  of a dog  (\textit{perceptual access}) and 
 correspondingly reasons and believes that there should be a dog nearby (\textit{rationally perceived knowledge  
 transformed into a belief}). 
This belief  makes her anxious (\textit{stimulated emotion}).%
\end{example}
This example illustrates how general preferences alongside beliefs may directly stimulate emotions in rational agents. 
Note that in this case, if Grace was not afraid of dogs, the belief that a dog is nearby by itself would not stimulate 
particular emotions in her.%
\medskip

\noindent
\textbf{Elements that are influenced by emotions:} 
The following example illustrates how emotions of rational agents may influence their previous goals.%
\begin{example}
\label{example:emotions_to_change_goals}
Hailey has planned to go to a party tonight (\textit{original goal}). 
In the afternoon, Hailey receives bad news that make her deeply sad (\textit{emotion}). 
As a consequence, she decides not to go to the party anymore (\textit{change in the goal due to the emotions}).%
\end{example}
This example shows that emotions may change the previous goals of rational agents. 
Although Hailey's general preference may be to participate in such parties, 
and she may in general be an extrovert with high levels of conscientiousness (i.e., 
self-discipline and tendency to follow her schedules), 
due to the depth of her sadness she may make a goal (i.e., skipping the party) that contradicts her initial goal that corresponded 
to her general preference, personality traits, etc.%
\medskip

\noindent
\textbf{Importance of personalising the perceptual access and rational reasoning processes: }
The first example below illustrates the importance of personalising the perception procedure of various rational agents. 
In the given example the tourist is the observed agent and the tour leader and the tourist's close friend are observer agents. 
The second example below demonstrates the importance of decomposing the process that yields the rationally perceived knowledge 
into the sub-processes that are explained in \autoref{subsec:observation-reasoning}.%

\begin{example}
\label{example:personalized_perception}
Suppose that a tourist tells her tour leader that she has already been to the historical city centre 
of the city they are visiting. 
The tour leader may suppose that the tourist has a perfect knowledge of the real-life data, 
including the church, old building of the City Hall, and all the souvenir shops 
(\emph{general world knowledge of the observed agent according to the observer agent}), 
while in her previous visit the tourist has overlooked the old building of the City Hall 
because souvenir shops appealed more to her. 
Then the general world knowledge considered by the tour leader for the tourist is inaccurate. 
Now suppose that the tourist tells her close friend (\emph{an observer agent who is aware of the personalised 
perception of the observed agent}) that she has once been to the historical city centre. The friend assumes - knowing the personalised perception procedure of the tourist - that she has overlooked the old building of the City Hall. 
\end{example}

\begin{example}
\label{example:decomposing_rational_perception}
Brian, Charlie, and Diana are inside a shopping mall. 
Although they cannot see the outside, before they entered the shopping mall it was sunny. 
Someone soaked in water enters the shopping mall.  
Brian does not notice this person (\emph{no updated perceptual access}) and thus, 
Brian keeps the \emph{belief} that outside is sunny.   
Charlie and Diana notice this person (\textit{updated perceptual access}). 
Charlie reasons and accordingly \emph{believes} that it must be raining now, 
while Diana reasons that this person has fallen into a ditch (\textit{different [personalised] rational reasoning}) 
and keeps the \emph{belief} that it is still sunny outside.
\end{example}
Note that in the above example, if an observer agent infers about the beliefs of Brian, Charlie, and Diana 
(all as observed agents), if their personalised perceptual access and rational reasoning processes are excluded,   
the observer agent may infer that all three observed agents believe that it is now raining outside.%
\medskip

\noindent
\textbf{Elements that bias the generation of beliefs:} 
The two following examples show how generation of beliefs of rational agents may be biased by their emotions and goals, respectively. 
In the first example Igor and Jane and in the second example Kevin and his father are the observed agents.%

\begin{example}
\label{example:biased_beliefs_by_emotions}
Igor and Jane are having a walk together. 
While walking, they both see a dog (\textit{real-life data}, which after perceptual access and rational reasoning, 
for both observed agents results in the belief that there is a dog nearby). 
Since Igor is afraid of dogs (\textit{general preference}), he feels afraid (\textit{emotion resulting from emotion trigger 1}), 
and starts to believe that the dog might harm him (\textit{belief biased by emotion}).  
Jane, however, does not feel any fear and thus believes there is no threat from the dog.  
\end{example}
In the above example although both observed agents initially had the same belief (i.e., there is a dog nearby) 
one of them develops a biased new belief about the dog because of his triggered emotions.%

\begin{example}
\label{example:biased_beliefs_by_goals}
Kevin is a football fan (\textit{general preference}). 
The team he supports is currently in the second place in the championship.
When all evidences are studied by an objective analyst, they conclude that - although not impossible yet - the chances that Kevin's favourite team wins the championship are very small (\textit{unbiased belief}).  
Since Kevin wants his team to win (Kevin's \textit{goal}), he believes that his team will win  (\textit{belief biased by a goal}). 
Now suppose that Kevin's father has the same general preference and goal as Kevin, 
while he has a much lower level of conscientiousness (a personality trait that implies Kevin's father 
is generally less stubborn and more flexible with regards to situations). 
Consequently, Kevin's father develops a much less strong belief than his son about their team winning the championship.
\end{example}
The above example illustrates that goals can bias the beliefs of rational agents, whereas personality 
traits - while not generating a bias by themselves - may regulate (boost or hinder) this influence.%
\medskip

\noindent
\textbf{Inverse inference of emotions from actions:} 
The following two examples illustrate the process of inverse inference of emotions based on the observed actions 
of rational agents, 
which is based on updating the belief and goal of the observed agent (due to the influence of emotions), respectively. 
The observer agent is specified by the first-person pronouns and the other person in the examples is the observed agent. %

\begin{example}
\label{example:inverse_inference_belief}
I see Anabel and smile at her. Anabel and I are friends so I expect Anabel to smile back (\textit{expected action}). 
Anabel, however, looks unfriendly instead and turns her back at me (\textit{observed action}).
Thus, I conclude that Anabel may feel negative emotions about me at the moment (\textit{inferred emotion}). 
\end{example}
In the above example, the observer agent initially estimates the belief of the observed agent to be ``we are good to each other''. 
The observed action of the observed agent, however, implies that this belief is wrong. 
Therefore, the observer agent updates the belief of the observed agent to ``Anabel is not good to me'' and deduces that she has negative emotions. 
Now in order for the observer agent to infer precisely about the emotion of the observed agent (e.g., whether
Anabel is angry or sad), the observer agent should be aware of the belief(s) and goal(s) of the observed agent 
in their previous interaction(s).%

Suppose that this morning Anabel showed me a picture of a dress. 
Her goal was to wear it for her sister's wedding (Anabel's goal in the previous interaction). 
I said the dress may not look nice on her. Anabel believed that I was being mean to her (Anabel's belief in the previous interaction). 
This made her feel \textbf{angry} at me, and thus in our next interaction she updated her belief  
from ``we are good to each other'' to ``we are not friends anymore''.%

\begin{example}
\label{example:inverse_inference_goal}
 
Last week I had a good discussion with Lewis about our projects.  
While walking on the campus, I come across Lewis. 
I know he has a vacancy for his new project. 
I tell to Lewis that I am seeking a new project to join.  
I suppose that Lewis has developed the goal of establishing a collaboration with me after our previous talk, 
so I expect him to invite me to join his new project (\textit{expected action}). 
Instead, Lewis wishes me good luck and leaves (\textit{observed action}). 
Clearly my inference about Lewis' goal was wrong. 
\end{example}
In this example, the observer agent expects a particular action from the observed agent 
based on inferring the observed agent's belief to be ``the observer agent has very high qualifications''. 
Although this belief is correctly deduced, the goal is wrongly inferred, because the observer agent has 
not considered the beliefs and goals of the observer agent during their last interaction, i.e., 
their talk (Lewis believed that he should always be better than his employees and while talking 
his goal was to show he knows the best),   
and thus the emotions that were triggered (Lewis feels threatened by the qualifications of the observer agent 
and thus develops the goal to avoid collaborating with him).%

\section{Online survey for assessment of the cognitive models based on real-life data}
\label{app:survey}

\begin{table}
\centering
\begin{tabular}{|ccc|c|}
\hline
\multicolumn{3}{|c|}{\textbf{Linguistic term}} & \multirow{2}{*}{\textbf{\begin{tabular}[c]{@{}c@{}}Numerical value\\ (Realization of the concept)\end{tabular}}} \\ \cline{1-3}
\multicolumn{1}{|c|}{\textbf{Belief}} & \multicolumn{1}{c|}{\textbf{Goal}} & \textbf{Emotion} &  \\ \hline
\multicolumn{1}{|c|}{Heavy rain} & \multicolumn{1}{c|}{I do not want it at all} & Very unhappy & -1 \\ \hline
\multicolumn{1}{|c|}{Light rain} & \multicolumn{1}{c|}{I do not want to do it} & Unhappy & -0.5 \\ \hline
\multicolumn{1}{|c|}{I do not know} & \multicolumn{1}{c|}{I have no preference} & Nothing & 0 \\ \hline
\multicolumn{1}{|c|}{Partially sunny} & \multicolumn{1}{c|}{I want to do it} & Happy & 0.5 \\ \hline
\multicolumn{1}{|c|}{Sunny} & \multicolumn{1}{c|}{I want it a lot} & Very happy & 1 \\ \hline
\end{tabular}
\caption{Linguistic terms and their corresponding numerical values used for the sliding bars. }
\label{tbl:linguistic-terms-bge}
\end{table}

Initially, participants are asked to categorise $17$ predefined outdoor activities based on the 
degree (e.g., dislike moderately) they have a preference for the activities. 
Overall, $7$ linguistic terms are considered to express the participant's degree of preference. 
Every scenario in the survey requests the participant to consider a real-life situation 
where the participant wants to do one of the specified outdoor activities. 
An introductory statement specified the inputs of that scenario for the user 
(e.g., the weather forecast as the rationally perceived knowledge and the accuracy of the source of 
the weather forecast as the general world knowledge).  
%
%
Next, the participant is asked about their expected outcome (i.e., belief), goal, and emotion in the given scenario. 
The answers can be provided via a sliding scale that varies from a linguistic term that corresponds to the value $-1$ 
to another linguistic term that corresponds to the value $1$, with totally $5$ linguistic terms (see \autoref{tbl:linguistic-terms-bge}). 
A sample of the survey is available online via \url{https://drive.google.com/file/d/16ig0OIAf0XCvac6HS02gWVKg-Ug4dHvo/view?usp=sharing}.%

\end{document}